\documentclass[preprint,aps]{revtex4-1}

\usepackage{graphicx}
\usepackage{color}
 \usepackage{natbib}
 \usepackage{mciteplus}
\usepackage{amsmath}
\usepackage{amssymb}
\usepackage{latexsym}

\newcommand{\PR}[1]{{\color{black} #1}}

\begin{document}
%
%
\title{Beyond the GW approximation: combining correlation channels}
\author{$^{1,2,4}$Pina Romaniello, $^{3,4}$Friedhelm Bechstedt, and $^{2,4}$Lucia Reining}
\affiliation{$^1$Laboratoire de Physique Th\'{e}orique-IRSAMC, CNRS, Universit\'{e} Paul Sabatier, F-31062 Toulouse Cedex, France} 
\affiliation{$^2$Laboratoire des Solides Irradi\'{e}s UMR 7642, CNRS-CEA/DSM, \'{E}cole Polytechnique, F-91128 Palaiseau, France}
\affiliation{$^3$Institut f\"{u}r Festk\"{o}rpertheorie und-Optik, Friedrich-Schiller-Universit\"{a}t, Max-Wien-Platz 1, 07743 Jena, Germany}
\affiliation{$^4$European Theoretical Spectroscopy Facility (ETSF)}
\date{\today} 
\begin{abstract}
In many-body perturbation theory (MBPT) the self-energy $\Sigma=iGW\Gamma$  plays the key role since it contains all the many body effects of the system. The exact self-energy is not known; as first approximation one can set the vertex function $\Gamma$ to unity which leads to the $GW$ approximation. \PR{The latter properly describes the high-density regime, where screening is important}; in the low-density regime, instead, other approximations are proposed, such as the T matrix, which describes multiple scattering between two particles. 

\PR{Here we combine the two approaches. Starting from the fundamental equations of MBPT we show how one can derive the T-matrix approximation to the self-energy in a common framework with $GW$}. This allows us to elucidate several aspects of this formulation, including the origin of, and link between, the electron-hole and the particle-particle T matrix, the derivation of a screened T matrix, and the conversion of the T matrix into a vertex correction. The exactly solvable Hubbard molecule is used for illustration.
\end{abstract}
\maketitle 
\section{Introduction}
The $GW$ approximation (GWA) to the electron self-energy $\Sigma$ \cite{hedin65} is nowadays the method of choice for band-structure \cite{Aulbur,Gunnarsson} and photoemission calculations (see e.g. Ref.\ \cite{gatti-VO2,gatti-V2O3,Kotani07,Kotani04}). In the general expression $\Sigma=iGW\Gamma$ it approximates the vertex function $\Gamma\approx 1$, and keeps  only the 1-particle Green's function $G$ and the screened Coulomb interaction $W$. In the high-density regime, where screening is important, the \textit{GWA} works reasonably well; in the low-density limit, instead, where the quantum nature of the system dominates, $GW$ shows some failures \cite{vonBarth,Stan,Godby_SI,Bruneval_PRL09,Aryasetiawan_Tmatrix,Pina09}. One hence needs to go beyond $GW$. \PR{Iterating Hedin's equations further seems the obvious thing to do, but this is technically difficult and there is no guarantee that results will quickly improve. To overcome the problems of finite-order corrections to the self-energy one might then use the strategy of creating an infinite number of diagrams via a Dyson equation for the Green's function. This is what is done in GW, where the Green's function is determined to infinite order in $W$ using a Dyson equation with a kernel (the self-energy) that is only linear in $W$. A similar strategy can be used for the self-energy itself by introducing the scattering T matrix \cite{Galitskii,Kanamori,fetterwal,kadanoffbaym_PR61} that yields the self-energy as $\Sigma=GT$. In the so-called Bethe-Goldstone approximation, originally introduced in the nuclear many-body problem \cite{bethegoldstone} for the two-particle Green's function, $T$ describes multiple scattering between two particles (two electrons or two holes) or an electron and a hole. The approximation is justified in the limit of low density of electrons or holes, i.e. close to completely filled or completely empty bands.

T-matrix approaches have been extensively used in the context of Hubbard models \cite{Cini_SSC86,Verdozzi_PRL95,Verdozzi_PRL,Verdozzi_PRB,Godby_PRB98} and the results have confirmed  that this approximation is very good at low-electron density, precisely where $GW$ fails, but is not superiour to the GWA at half filling where the correlation gap is not well reproduced \cite{Verdozzi_PRB}. However in general none of the three possibilities (GWA, particle-particle or electron-hole T matrix) will give an exhaustive description. This reflects the dilemma of how to decide which two-particle correlation to privilege in the description of a (at least) three-particle problem. It suggests to work with
combinations. There are several ways to combine different correlation channels, and care must be taken to prevent double counting of low-order terms \cite{Bickers_2,Bickers_3}. The "fluctuating exchange" (FLEX) approximation by Bickers \textit{at al.} \cite{Bickers_PRL,Bickers_AOP} starts from the second Born self-energy, i. e. the exact self-energy to second order in the Coulomb potential, and then sums all
contributions, starting from the third order, in each channel separately (i.e. there are no mixed diagrams). It contains hence the GWA and its exchange counterpart to all orders, plus, starting from third order, all T-matrix particle-particle diagrams, and all  T-matrix
electron-hole diagrams. In particular for the Hubbard model with its local interaction, where the exchange contribution simply
leads to a spin-dependence of the interaction, calculations remain quite efficient. The Bethe-Goldstone T-matrix and GW approximations can be regarded as an approximation to FLEX where only one channel is taken into account. Beyond the summation of independent channels two-step FLEX approaches have been proposed, where one channel enters the calculation of a second channel through
an effective screening \cite{Liebsch,Katsnelson_JPCM,Katsnelson_EPJB}. One can also couple particle-particle and electron-hole channels on an equal footing, which then leads to the rather involved parquet theory \cite{smith,Bickers_2}. In the same spirit in Ref\ \cite{Robert_PRB06} a variational functional of the Green's function and the two-particle scattering vertex (which is a T matrix as defined in \cite{Strinati}) has been proposed. Its systematic construction yields the particle-particle T-matrix approach as simple approximation, and adds electron-hole diagrams at a higher level of perturbation theory. 

In this work we propose an alternative way to derive the coupling of GW and T-matrix channels on an equal footing starting from exact many-body equations. Attempts to  go beyond GW by summing the GW self-energy to screened versions of the T-matrix self-energy are found in literature \cite{Aryasetiawan_Tmatrix,Chulkov_PRL,Chulkov_PRB,Chulkov_PRB06,Nechaev_PRB06,Nechaev_PRB08,Monnich_PRB06}. However only putting the different approximations on the same footing one can get unambiguous corrections to GW from the T matrix. These screened T-matrices indeed require appropriate double counting corrections to keep the second-order terms exact and to avoid negative spectral functions.}
Therefore in the following we present a unified framework that links $GW$, $GW\Gamma$ and T matrix. This allows us to address several questions, in particular, \textit{What is the origin of, and link between, the particle-particle and the electron-hole contributions to the T matrix? How do we get a screened version of the T matrix? How do we translate the physical content of the T matrix into a vertex correction?} These questions will be answered in Sec.\ \ref{Section_II}. In Sec. \ref{Applications} we will then apply the T matrix to the Hubbard molecule at 1/4 and 1/2 filling. This system allows us to compare the T matrix, GW, and the exact results, hence to illustrate the performances of the different approximations. Conclusions are given in Sec. \ref{Conclusions}. 
\section{A Unified framework\label{Section_II}}
In order to use a common language for $GW\Gamma$ and T matrix, we start from the following exact expression for the self-energy:
\begin{equation}
\Sigma(11^\prime)=-iv_c(1^+2)G_2(12;32^+)G^{-1}(31^\prime),
\label{Eqn:SE_exact}
\end{equation}
\PR{where $(1)=\displaystyle  (r_1,\sigma_1,t_1) $, $(1^+)=(r_1,\sigma_1,t^+_1)$ with $t_1^+=t_1+\delta$ ($\delta\rightarrow 0^+$) describe space, spin and time coordinates, and integration over indices
not present on the left is implicit throughout the paper}. By adding a perturbing potential $U_{ext}$ and using the relation $\left.G_2(12;32^+;[U_{ext}])\right|_{U_{ext}=0}=G(13)G(22^+)-\left.\frac{\delta G(13)}{\delta U_{ext}(2)}\right|_{U_{ext}=0}$ \cite{martsch59}, (\ref{Eqn:SE_exact}) can be written as
\begin{equation}
\Sigma(11^\prime)=v_H(1)\delta(11^\prime)-iv_c(1^+2)G(13)\left.\frac{\delta G^{-1}(31')}{\delta U_{ext}(2)}\right|_{U_{ext}=0},
\label{Eqn:Self-energy_0}
\end{equation}
where $\frac{\delta G}{\delta U_{ext}}=-G\frac{\delta G^{-1}}{\delta U_{ext}}G$ is used. Here $v_H(1)=-iv(1^+2)G(22^+)$ is the Hartree potential and the second term on the right-hand side defines the exchange-correlation contribution to the self-energy, $\Sigma_{xc}$. With the help of the Dyson equation for $G$, Eq.\ (\ref{Eqn:Self-energy_0}) can be further rearranged as
\begin{equation}
\Sigma(11^\prime)=v_H(1)\delta(11^\prime)+\Sigma_x(11^\prime)+iv_c(1^+2)G(13)\Xi(35;1^\prime 4)L(42;52^+),
\label{Eqn:Self-energy}
\end{equation}
with $\Sigma_x(11')=iv(1^+1')G(11')$, $\Xi(35;1^\prime 4)=\frac{\delta\Sigma(31^\prime)}{\delta G(45)}$ the effective interaction, and $L(42;52)=\left.\frac{\delta G(45)}{\delta U_{ext}(2)}\right|_{U_{ext}=0}$ the time-ordered "response" of the system to an external perturbation $U_{ext}$. This way to write the self-energy directly displays the physics behind it, i.e. the description of a particle interacting with the system: the particle can scatter against the density of the system (Hartree term), it can exchange with another particle of the system (exchange term), it can do something to the system (last term), i.e. it can have an effective interaction with the system ($\Xi$), the system responds ($L$), and the particle feels this response through the Coulomb interaction ($v_c$).

There are two essential ingredients in Eq.\ (\ref{Eqn:Self-energy}): the effective interaction $\Xi(35;1^\prime 4)$, and the response of the system $L(42;52)$. Combining approximations to $\Xi$ and to $L$, various approximations to the self-energy can be created. In situations where the screening is important one should make an effort to obtain a good $L$, whereas in situations where the quantum nature of the interaction is important \footnote{The atomic limit of the Hubbard molecule is an example where the correlation part of the interaction is crucial; see later} one would concentrate on $\Xi$, although  $L$ and $\Xi$ are of course in principle linked through the Bethe-Salpeter equation \cite{Strinati} and one might wish to keep them approximately consistent. %
\subsection{How to get \PR{$GW$}?}
Neglecting the variation of $\Sigma_{xc}$ in $\Xi$, i.e. keeping only the classical interaction $v_c$, one obtains $\Sigma_{xc}(11') = \Sigma_x + iv_c(12)G(11')v_c(1'4)\chi(42)$, with $\chi(42)=-iL(42;42)$ \PR{the time-ordered response function}. Hence one gets a screening contribution with respect to $\Sigma_x$: this is the GW form, with $W=v_c+v_c\chi v_c$. At this stage it has not been specified yet how to calculate the screening: different approximations to the screening will give the various GW flavours (e.g. GW$^{RPA}$ and beyond \footnote{In this case the nature of the screening would be always test-charge-test-charge \cite{Bruneval_PRL05}}). If one keeps an approximate $\Sigma_{xc}$ in $\Xi$ one goes beyond GW and includes vertex corrections. For example, approximating $\Sigma_{xc}$ by \PR{the exchange-correlation potential of DFT, $v_{xc}$,} one gets $\Sigma_{xc}(11') = \Sigma_x + iv_c(12)G(11')\left [v_c(1'4)+f_{xc}(1'4)\right ]\chi(42)$ \PR{, where $f_{xc}=\frac{\delta v_{xc}}{\delta\rho}$}; this leads to 
$\Sigma_{xc}=iGW\Gamma$ with an approximate vertex function $\Gamma=1+f_{xc}P$, where $P=iGG\Gamma$ is the irreducible polarizability and we used $\chi=P+Pv_c\chi$ \footnote{In this case one can write $\Sigma_{xc}=iG\tilde{W}$ where $\tilde W$ is a test charge-test electron (TC-TE) screening \cite{Bruneval_PRL05}. The $f_{xc}$ that appears makes the one-electron case exact \cite{Pina09}.}.

\subsection{How to get the T matrix?}
One could also use the rough approximation  $L(42;52)=- 
G(47)\frac{\delta G^{-1}(78)}{\delta U_{ext}( 2)}G(85)\approx G(4 2)G(25)$ but concentrate on a clever approximation for $\Xi$. This modifies the exact self-energy (\ref{Eqn:Self-energy}) as 
\begin{equation}
 \Sigma(11') \approx v_H(1)\delta(11') + \Sigma_x(11') + i v_c(1 2) G(13)\left[ \frac {\delta \Sigma(31')}
{\delta G(45)}G(4 2)G(25)\right].
\label{Eqn:S_O}
\end{equation}
\subsubsection{An effective 4-point interaction $O$}
It still remains to find an appropriate approximation for the functional derivative on the right-hand side of Eq.\ (\ref{Eqn:S_O}). One can introduce an effective 4-point interaction $O$ such that, similar to GW, 
\begin{equation}
 \Sigma(11') = G(4 2) O(1 2;1'4)
 \label{Eqn:SE_BG1}
\end{equation}
\PR{Note that Eq.\ (\ref{Eqn:SE_BG1}) is closely related to the expression of the self-energy within the T-matrix approximation as given e.g. by Kadanoff and Baym (see Eq.\ (56) in Ref. \cite{kadanoffbaym_PR61}), which is the goal of this derivation. However at this stage $O$ is not yet the T matrix.} Since $G(4 2) O(1 2;1'4)$ cannot be inverted to find $O$, several choices of $O$ make the correct $\Sigma$ \footnote{This means that $O$ is only determined up to changes $\Delta O$ that fulfill the condition $\int d3d4 G(34)\Delta O(14;23)=0$.}. First note that in (\ref{Eqn:S_O}) there are direct and exchange terms. Therefore it is convenient to divide the self-energy as $\Sigma=\Sigma_1+\Sigma_2$ and, consequently, $O=O_1+O_2$ with
\begin{eqnarray}
 O_1(1 2;1'4) &=& -iv_c(12)\delta(11')\delta(4 2)  +iv_c(1 2) G(13)\left[ \frac {\delta \Sigma_1(31')}
{\delta G(45)}G(25)\right],
\label{Eqn:O_pp_1}\\
 O_2(1 2;1'4) &=& iv_c(1 2)\delta(21')\delta(41) +iv_c(1 2) G(13)\left[ \frac {\delta \Sigma_2(31')}
{\delta G(45)}G( 25)\right].
\label{Eqn:O_pp_2}
\end{eqnarray}
\PR{This decomposition of $O$ allows us to find two interaction channels as in the T-matrix self-energy with a direct term (here given by $O_1$) and an exchange term (here given by $O_2$) (see e.g. Eq.\ (13.23) of Ref.\ \cite{kadanoffbaym}).}

The solutions (\ref{Eqn:O_pp_1}) and (\ref{Eqn:O_pp_2}) are not unique; one could equally have written 
\begin{equation}
 \Sigma(11') = G(25) O(15;1'2) := G(25)\left [O_1(15;1'2)+O_2(15;1'2)\right ]:=\Sigma_1(11') +\Sigma_2(11') 
\label{Eqn:SE_BG2},
\end{equation}
with 
\begin{eqnarray}
O_1(15;1'2)&=& -iv_c(1' 2)\delta(11')\delta(5 2)+i v_c(1 2) G(13) \left[ \frac {\delta \Sigma_1(31')}
{\delta G(45)}G(42)\right],
\label{Eqn:O_eh_1}\\
O_2(15;1'2)&=& iv_c(1' 2)\delta(12)\delta(5 1')+i v_c(1 2) G(13) \left[ \frac {\delta \Sigma_2(31')}
{\delta G(45)}G(42)\right].
\label{Eqn:O_eh_2}
\end{eqnarray}
These two decompositions of the self-energy are equivalent, i.e. they give the same self-energy if the exact $\Sigma$ is used. 
\subsubsection{A Dyson equation for O: the particle-particle and electron-hole T matrix}
At this level we do not have yet a closed expression for $O$, but Eq.\ (\ref{Eqn:SE_BG1}) suggests an approximation to the functional derivative, in analogy with what one usually does in the framework of Bethe-Salpeter calculations based on GW \cite{Hanke,Strinati_PRB,Strinati_PRL} \PR{\footnote{Also in the Bethe-Salpeter equation used for the calculation of the electron-hole excitations one has to approximate the kernel  $\Xi=\frac{\delta\Sigma}{\delta G}$. Using the GW approximation to the self-energy, i.e. $\Sigma^{GW}=iGW$, the kernel is approximated as $\Xi\approx W$, thus neglecting the term $\frac{\delta W}{\delta G}$ in the functional derivative of $\Sigma^{GW}$. In this respect the approximation presented in Eq.\ (\ref{Eqn:approximation}) is done in the same spirit as in the GW-BSE framework}}:

\begin{equation}
 \frac {\delta \Sigma_i(31')}{\delta G(45)}\approx O_i(3 5;1'4).
\label{Eqn:approximation}
\end{equation}
Note that, with (\ref{Eqn:S_O}) and (\ref{Eqn:approximation}), $O$ is only an approximation to the total interaction $\Xi$. Note also that approximation (\ref{Eqn:approximation}), together with Eq.\ (\ref{Eqn:SE_BG1}) and reference \footnotemark[4],  is along the same line of the approximations based on the Ward identity used, e.g., in Refs  \cite{Ng_1,Ng_2,Vignale_book, Asgari}. 

This approximation used in (\ref{Eqn:O_pp_1}) and (\ref{Eqn:O_pp_2}) allows one to determine $O$ from an integral equation 
\begin{eqnarray}
 O^{pp}_1(1 2;1'4) &=& -iv_c(12)\delta(11')\delta(4 2)+i v_c(1 2) G(13)G(25)O^{pp}_1(3 5;1'4), 
\label{Eqn:T-matrix_pp_1}\\
O^{pp}_2(1 2;1'4) &=&iv_c(12)\delta(21')\delta(4 1)+i v_c(1 2) G(13)G(25)O^{pp}_2(3 5;1'4).
\label{Eqn:T-matrix_pp_2}
\end{eqnarray}
Here the subscript $pp$ indicates the particle-particle nature of the interaction $O$ that comes from the time ordering of the kernel $GG$ (see App.\ \ref{Appendix1}).
Note that the zeroth order term in Eq.\ (\ref{Eqn:T-matrix_pp_2}) has simply a minus sign and the indices 4 and 1', which, moreover, are external indices in $O^{pp}_i$, exchanged with respect to Eq.\ (\ref{Eqn:T-matrix_pp_1}) for $O^{pp}_1$; therefore one can relate $O_2^{pp}$ to $O^{pp}_1$ as
\begin{equation}
O^{pp}_2(1 2;1'4)= -O^{pp}_1(1 2;41').
\label{Eqn:antisymm}
\end{equation}
$O^{pp}=O^{pp}_1+O^{pp}_2$ with $O^{pp}_1$ and $O^{pp}_2$ given by (\ref{Eqn:T-matrix_pp_1})-(\ref{Eqn:T-matrix_pp_2}) can be identified with the \textit{particle-particle} T matrix \cite{Galitskii,kadanoffbaym}. \PR{Using relation (\ref{Eqn:antisymm}) one can verify that the $pp$ T matrix meets the cross relation $O^{pp}(12;1'4)=-O^{pp}(12;41')$ arising from the Pauli principle. Moreover, one can also verify that $O^{pp}(12;1'4)=O^{pp}(1'4;12)$, i.e. the T matrix remains the same for a scattering process reversed. }

Equivalently, using the approximation (\ref{Eqn:approximation}) for $O$  in Eqs (\ref{Eqn:O_eh_1})-(\ref{Eqn:O_eh_2}), one also obtains Bethe-Salpeter-like equations
\begin{eqnarray}
\label{eq:o1}
 O^{eh}_1(15;1'2)& =& -iv_c(1' 2)\delta(11')\delta(5 2)+i v_c(12) G(13)G(42) O^{eh}_1(3 5;1'4),
 \label{Eqn:T-matrix_eh_1}\\
O^{eh}_2(15;1'2)& =& iv_c(1'2)\delta(12)\delta(51')+i v_c(12) G(13)G(42) O^{eh}_2(3 5;1'4),
\label{Eqn:T-matrix_eh_2}
\end{eqnarray} 
where the subscript $eh$ indicates the \textit{electron-hole} nature of the T matrix.
Note that, similar to the particle-particle T matrix, the zeroth order term in $O^{eh}_2$ has simply a minus sign and the indices 1 and 5 exchanged with respect to $O^{eh}_1$. However, the fact that 1 and 5 are not external indices prevents from relating $O_2^{pp}$ to $O^{pp}_1$, unlike in the case of the pp T matrix \footnote{One might wonder whether the eh T matrix (\ref{Eqn:T-matrix_eh_1})-(\ref{Eqn:T-matrix_eh_2}) is the same as the one which is obtained using the T matrix derived by Strinati in Ref.\ \cite{Strinati} when only ladder diagrams are used (i.e. $\frac{\delta\Sigma(11')}{\delta G(2'2)}\approx iv_c(11')\delta(12')\delta(1'2)$). One can verify that this is not the case since there is only an exchange-like term similar to $O^{eh}_2$, and a different integration of variables.}. \PR{This is because the Pauli principle does not apply. The symmetry with respect to a reversed scattering process, instead, holds.} 

\begin{figure}[htbp]
\begin{center}
\caption{Diagrams corresponding to the self-energy obtained with the second iteration of the particle-particle T matrix ($\Sigma^{pp,(2)}_1$  and $\Sigma^{pp,(2)}_2$)and electron-hole T matrix ($\Sigma^{eh,(2)}_1$  and $\Sigma^{eh,(2)}_2$).}
\vspace{4em}
\includegraphics[width=\textwidth]{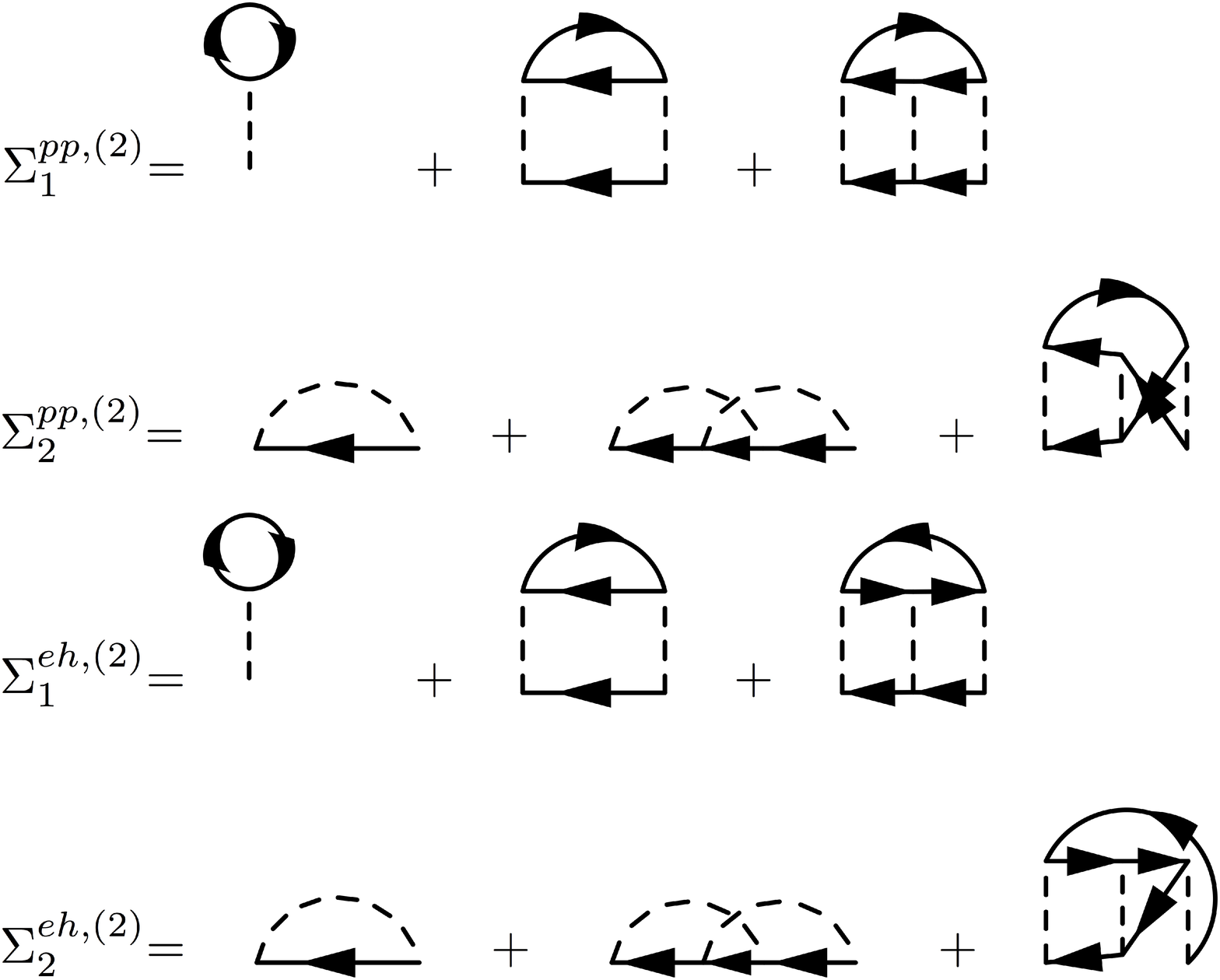}
\label{figure1}
\end{center}
\end{figure}

Because of the approximation (\ref{Eqn:approximation}), the pp T matrix and the eh T matrix do not in general give the same self-energy anymore. In the first iteration of (\ref{Eqn:T-matrix_pp_1})-(\ref{Eqn:T-matrix_pp_2}) and (\ref{Eqn:T-matrix_eh_1})-(\ref{Eqn:T-matrix_eh_2}) the equality still holds, with 
 \begin{eqnarray}
 \Sigma^{pp,(1)}(11')=\Sigma^{eh,(1)}(11')=T_0(11')&=&v_H(1)\delta(11')+\Sigma_x(11') \nonumber\\
 &+& v_c(12)v_c(1'5)G(11')G(52)G(25)\nonumber\\
  &-& v_c(12)v_c(1'5)G(21')G(52)G(15),
  \end{eqnarray}
 \PR{which is the second Born approximation.}
 With the second iteration differences appear (see Fig. \ref{figure1} ). Indeed one obtains
\begin{eqnarray}
\Sigma^{pp,(2)}(11')&=&T_0(11')+T_1(11'),\\
\Sigma^{eh,(2)}(11')& =&T_0(11')+T_2(11')
\end{eqnarray}
with
\begin{eqnarray}
T_1(11')&=& iv_c(12)v_c(35)v_c(1'4)G(13)G(31')G(54)G(42)G(25)\nonumber\\
&-&iv_c(12)v_c(35)v_c(1'4)G(13)G(34)G(51')G(42)G(25),\\
T_2(11')&= &iv_c(12)v_c(34)v_c(1'5)G(13)G(31')G(54)G(42)G(25)\nonumber\\
&-&iv_c(12)v_c(34)v_c(1'5)G(13)G(35)G(54)G(42)G(21').
\end{eqnarray}
If one considers also the term $\delta O/\delta G$ in the iteration (in (\ref{Eqn:O_pp_1}) and (\ref{Eqn:O_pp_2}) or (\ref{Eqn:O_eh_1}) and (\ref{Eqn:O_eh_2})), then the equality of the self-energy via a $pp$ or a $eh$ channel is re-stablished, i.e.
\begin{equation}
\Sigma^{pp,(2)}_{full}(11') = \Sigma^{eh,(2)}_{full}(11') = T_0(11')+T_1(11')+T_2(11')+T_3(11')
\label{Eqn:S_O_2}
\end{equation}
with
\begin{eqnarray}
T_3(11')&=& iv_c(12)v_c(35)v_c(1'4)G(13)G(32)G(21')G(45)G(54)\nonumber\\
&-&iv_c(12)v_c(35)v_c(1'4)G(13)G(32)G(24)G(45)G(51').
\end{eqnarray}
Therefore in case of the pp T matrix the two terms $T_2$ and $T_3$ come from the functional derivative $\delta O^{pp,(1)}/\delta G$, whereas in case of the eh T matrix the two terms $T_1$ and $T_3$ are the ones which are due to $\delta O^{eh,(1)}/\delta G$. We also note that $\delta O^{(0)}/\delta G = 0$, i.e. the functional derivative does not contribute to the first iteration, which explains why the pp and eh self-energy are the same after the first iteration. One can also rewrite Eq.\ (\ref{Eqn:S_O_2}) as
\begin{equation}
\Sigma^{(2)}_{full}=\left(\Sigma^{pp,(2)}+\Sigma^{eh,(2)}\right)-T_0+T_3,
\label{Eqn:Functional_derivative_1}
\end{equation}
or alternatively, for example, as
\begin{equation}
\Sigma^{(2)}_{full}=\frac{1}{2}\left(\Sigma^{pp,(2)}+\Sigma^{eh,(2)}\right)+\frac{1}{2}\left(T_1+T_2\right)+T_3.
\label{Eqn:Functional_derivative_2}
\end{equation}

This shows that one might try to sum the pp and eh self-energies to account for some terms of the functional derivatives $\delta O/\delta G$, but such a sum is not obvious. For example, \textit{would it be worse to neglect (besides $T_3$) the term $-T_0$ or $(T_1+T_2)/2$?}  We will discuss this issue in Sec. \ref{Applications}.

\begin{figure}[htbp]
\begin{center}
\caption{Schematic representation of the physical contents of GW (a), pp T matrix (b), and eh T matrix for particles with collinear spins.}
\vspace{4em}
\includegraphics[width=\textwidth]{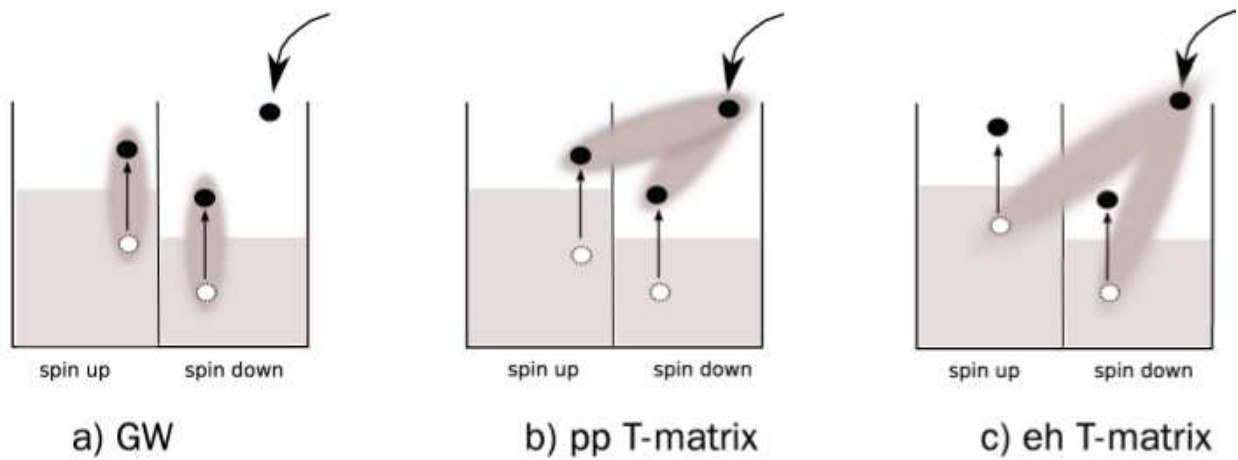}
\label{figure2}
\end{center}
\end{figure}

Based on (\ref{Eqn:Self-energy}), we can now directly compare the different approximations: in GW or an approximate GW$\Gamma$, one fixes an approximation for the functional derivative of the self-energy to a low-order approximation, but one tries to treat $L$ well. In the T-matrix approximation the latter is treated quite badly, whereas the former is kept, though approximatively, to infinite order in $v_c$ thanks to the Dyson-like equation for the effective interaction. Both approximations allow one to describe physical processes involving three particles: the particle which is added to the system and the electron-hole pair that it creates. Ideally one would propagate the three particles together, which is not numerically affordable. Therefore one chooses to propagate a pair and to treat the third particle in a kind of mean-field of the other two. This is illustrated in Fig.\ (\ref{figure2}): in GW one propagates together the electron-hole pair created by the additional particle, whereas in the T matrix one propagates together the additional electron (additional hole) and the excited electron (hole left behind) (pp T matrix) or the additional electron (additional hole) and the hole left behind in the electron excitation (excited electron) (eh T matrix). The choice is not obvious, therefore it is desirable to go beyond the simple scheme and to include terms coming from both the approaches. Before we have discussed pp and eh T matrix. Now we go towards a scheme which combines T matrix and GW. In order to do so we can see wether a Dyson equation for the T matrix can also be obtained with an improved approximation for $L$.
\subsubsection{Screened T matrix I \label{Screened_T_ok}}

Equations (\ref{Eqn:T-matrix_pp_1})-(\ref{Eqn:T-matrix_pp_2}) and (\ref{Eqn:T-matrix_eh_1})-(\ref{Eqn:T-matrix_eh_2}) define the equations for the T matrix. One can go beyond this approximation and include the Hartree potential or even a local part of the xc self-energy $\Sigma^{loc}_{xc}$ in the variation $\delta G^{-1}/\delta U_{ext}$. This yields $L(42;52)\approx G(47)\epsilon^{-1}(72)G(75)$ where $\epsilon^{-1} $ is a test charge-test charge screening function $\epsilon^{-1}=1+v\chi$ if only the Hartree part is included, otherwise at least a partially test charge-test electron screening $\epsilon^{-1}=1+\left[v+\frac{\delta\Sigma^{loc}_{xc}}{\delta\rho}\right]\chi$ \cite{Bruneval_PRL05}. This makes the expression much more, though not fully, consistent: now $\epsilon^{-1}$ 
 contains a large part of the derivative of the self-energy, that is
also considered in the effective interaction, and $L$ contains the screening
of the formerly independent propagators $GG$, that is itself based on the
two-particle correlation function. Eq.\ (\ref{Eqn:Self-energy}) then becomes 
\begin{equation}
 \Sigma(11') \approx v_H + \Sigma_x + iv_c(1 2) \epsilon^{-1}(72)G(13)\left[ \frac {\delta \Sigma(31')}
{\delta G(45)}G(47)G(75)\right].
\label{Eq:S_screened-O}
\end{equation}
Note that this is the same equation as (\ref{Eqn:S_O}), but now with $W=\epsilon^{-1}v_c$ replacing the bare Coulomb interaction under the integral. This leads to a screened matrix O:  $\Sigma(11') = G(42)O^{pp}_s (12;1'4)$, with $O^{pp}_s=O^{pp}_{s,1}+O^{pp}_{s,2}$  and
\begin{eqnarray}
 O^{pp}_{s,1}(12;1'4) &= &-iv_c(12)\delta(11')\delta(42)  +i W(12) G(13)G(25) O^{pp}_{s,1}(35;1'4),
 \label{Eq:T_screened_1}
 \\
O^{pp}_{s,2}(12;1'4)&=&=iv_c(12)\delta(14)\delta(1'2)  +i W(12) G(13)G(25) O^{pp}_{s,2}(35;1'4).
\label{Eq:T_screened_2}
\end{eqnarray}
In principle one could include also a nonlocal part of the xc self-energy, but the equations become more involved.  

As for the T matrix, also for the screened version one has $O^{pp}_{s,2}(12;1'4)=-O^{pp}_{s,1}(12;41')$.
Similarly,  one can derive the electron-hole screened T matrix, which looks like Eqs (\ref{Eqn:O_eh_1})-(\ref{Eqn:O_eh_2}) with $W$ replacing the bare Coulomb potential in the last term on the right-hand side. When $\epsilon^{-1}=1$, and hence $W=v_c$, this version of the screened T matrix reduces to the one of Ref.\ \cite{kadanoffbaym_PR61} \footnote{Since the screened Coulomb potential is not static, the time structure of the screened T matrix is more complicated than that of the T matrix. In particular one does not have anymore a strict particle-particle or electron-hole $L_0=-iGG$. Nevertheless we will keep this notation in order to make a link with the T matrix. }. Note that the Hartree and exchange parts remain unscreened, which is in net contrast with other versions of the screened T matrix reported in literature \cite{Aryasetiawan_Tmatrix,Chulkov_PRL,Chulkov_PRB,Chulkov_PRB06,Nechaev_PRB06,Nechaev_PRB08,Monnich_PRB06}.

It is interesting to take the screened T-matrix equation in its first iteration: in this case
the self-energy (particle-particle version) becomes 
\begin{eqnarray}
\Sigma^{pp,(1)}_{s,1}(11') &=& \delta(11')v_H (1) + iW (12)G(11')v_c(1'4)L_0(24),\\	
\Sigma^{pp,(1)}_{s,2}(11')&= &\Sigma_x(11')-W (12)G(14)v_c(41')G(21')G(42),
\end{eqnarray}
with $L_0(24)=-iG(24)G(42)$. The electron-hole screened T matrix produces the same self-energy as the particle-particle screened T matrix in its first iteration, as in case of the T-matrix approximation. \PR{Moreover the resulting self-energy is exact to second-order in the Coulomb interaction.}

\begin{figure}[htbp]
\begin{center}
\caption{Diagrams corresponding to the self-energy obtained with the first iteration of the screened T matrix. The diagrams, from left to right, represent the Hartree, GW, and second-order screened exchange (SOSEX) terms, respectively. Note that in the GW term we collapsed the two terms which for $\epsilon^{-1}=1$, i.e. $W=v_c$, reduce to the second and first diagrams of $\Sigma^{pp/eh}_1$ and $\Sigma^{pp/eh}_2$, respectively, of Fig.\ (\ref{figure1}).}
\vspace{4em}
\includegraphics[width=\textwidth]{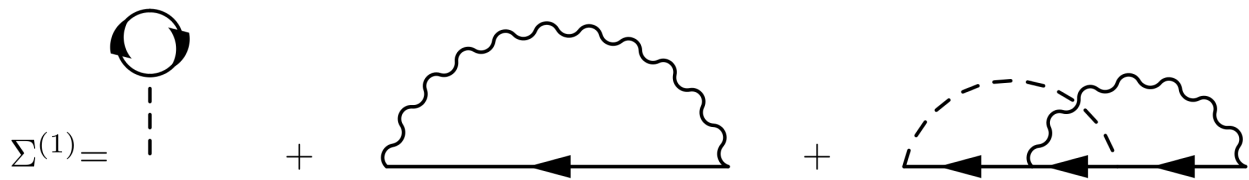}
\label{figure3}
\end{center}
\end{figure}

In the RPA, $W(12)v_c(1'4)L_0(24) = [v_c +v_cL_0v_c/(1-L_0)v_c]L_0v_c = W-v_c$. The sum of $\Sigma_1$ and $\Sigma_2$ yields hence GW, plus the last term of $\Sigma_2$. In Fig.\ \ref{figure3} we report the diagrammatic representation of this self-energy: the first two diagrams represent the Hartree and GW contributions, respectively, whereas the last one is a term corresponding to the second-order screened exchange (SOSEX). The latter contribution is becoming popular as correction to RPA in order to produce accurate results in the description of electronic correlation in atoms and solids \cite{SOSEX}. 

We hence can conclude that GW is contained in this screened T-matrix approach, which moreover contains promising higher-order terms. In literature other versions of the screened T matrix  \cite{Aryasetiawan_Tmatrix,Chulkov_PRL,Chulkov_PRB,Chulkov_PRB06,Nechaev_PRB06,Nechaev_PRB08,Monnich_PRB06} are proposed which are combined with the $GW$ approximation to get the total self-energy. However, since some of the terms in the T matrix are already contained in the GW approximation, care must be taken to avoid double counting. In our formulation, instead, the screened T matrix naturally contains GW; there is hence no need to add \textit{ad hoc} corrections.

\PR{Because of the appearance of both $v_c$ and $W$ the screened T-matrix approximation (\ref{Eq:T_screened_1})-(\ref{Eq:T_screened_2}) does not fulfill some symmetry conditions to be fully conserving, unlike self-consistent GW and (unscreened) T matrix \cite{kadanoffbaym_PR61}. For example, the momentum conservation law is violated.}
%
\subsubsection{Screened T matrix II \label{Screened_T_ko}}
Starting from (\ref{Eqn:Self-energy}), one can also write 
\begin{eqnarray}
\Sigma_{xc}(11') &= &\Sigma_{x}(11') +iv_c(12)G(13)\left[\frac{\delta v_H(3)\delta(31')}{\delta G(45)}L(42;52)+ \frac{\delta\Sigma_{xc}(31')}{\delta G(45)}L(42;52)\right]\nonumber\\
& =&\Sigma_{GW} +iv_c(12)G(13)\frac{\delta\Sigma_{xc}(31')}{\delta G(45)}L(42;52).	
\end{eqnarray}

 Using now again the approximation $L(42; 52) \approx G(47)\epsilon^{-1}(72)G(75)$ one obtains 
\begin{equation}
\Sigma_{xc}(11')\approx\Sigma_{GW} +i W(17)G(13)\frac{\delta\Sigma_{xc}(31')}{\delta G(45)}G(47)G(75).
\end{equation}

Note that here only the xc part of the self-energy appears. The ansatz will now use a screened xc matrix O: $\Sigma_{xc}(11') = G(42)O^{pp}_{sxc}(12; 1'4)$ (or equivalently $\Sigma_{xc}(11') = G(25)O^{eh}_{sxc}(15; 1'2)$); the structure of the resulting equation 
\begin{equation}
O^{pp}_{sxc}(12; 1'4) = iW (12)\delta(21')\delta(41) + iW(12)G(13) [O^{pp}_{sxc}(35; 1'4)G(25)]
\label{Eqn:Screened_T_ko}
\end{equation}
 is equal to the one determining the screened $O^{pp}_{s,2}$, Eq. (\ref{Eq:T_screened_2}), which for $\epsilon^{-1}=1$ reduces to $O^{pp}_2$. Similarly the structure of the electron-hole T-matrix $O^{eh}_{sxc}(15; 1'2)$ is equal to the one determining the screened $O^{eh}_{s,2}$, which for $\epsilon^{-1}=1$ reduces to $O^{eh}_2$. Note that this screened T matrix yields a self-energy corresponding to the exchange term only of Refs \cite{Aryasetiawan_Tmatrix,Chulkov_PRB}, where a screened T matrix is also used.
 
 Iterating this equation for $O^{pp}_{sxc}$ once, one gets the $GW\Gamma^{(1)}$ \cite{hedin65,Shirley_PRB96} approximation, namely: 
\begin{equation}
O^{pp,(1)}_{sxc} (12; 1'4) \approx iW (12)\delta(21')\delta (41) - W(12)G(14)W (41')G(21'),
\label{Eqn:O_xc_1}	
\end{equation}
from which $\Sigma_{xc}(11') = \Sigma^{GW}_{xc}-G(42)W(12)G(14)W (41')G(21')$. One can verify that the electron-hole screened T matrix produces the same self-energy as the particle-particle screened T matrix in its first iteration, as in case of the T-matrix approximation. It is, not surprisingly, also 
what one obtains by iterating Hedin's equations \cite{hedin65}: starting from $\Sigma_{xc} = iGW$ the equation for the irreducible vertex function reads
\begin{equation}
\Gamma(12;3) = \delta(12)\delta(13) + iW (21)G(16)G(72)\Gamma(67;3).
\end{equation}
To first order one gets hence $\Gamma(12;3) = \delta(12)\delta(13) + iW (21)G(13)G(32)$ and
\begin{equation}
\Sigma_{xc}(11') = iG(14)\Gamma(41';2)W (21) \approx \Sigma_{xc}^{GW}- G(14)W (1'4)G(42)G(21')W (21),
\end{equation}
which is exactly the xc self-energy obtained using (\ref{Eqn:O_xc_1}). There is instead no equivalent to (\ref{Eq:T_screened_1}) or its eh counterpart. However, in this second version of the screened T matrix, Hartree and exchange are not treated on the same level, since the latter is included in the definition of the T matrix, and the former not. This also explains why for $\epsilon^{-1}=1$ this screened T matrix does not reduce to the T matrix; in fact it reduces to only a part of it, i.e., to $O_2$. In Sec. \ref{Applications} we will show that this unbalance prevents this version of the screened T matrix to give the exact result for one electron in the atomic limit, unlike the screened T matrix given by Eqs (\ref{Eq:T_screened_1})-(\ref{Eq:T_screened_2}).
\subsubsection{Vertex corrections from T matrix}
Above we have shown that the pp screened T matrix (\ref{Eq:T_screened_1})-(\ref{Eq:T_screened_2}) (as well as the eh screened T matrix) contains GW plus extra terms beyond GW. We can therefore use the T matrix to formulate vertex corrections beyond GW. Starting from the self-energy $\Sigma=GO^{pp}_s=G[O^{pp}_{s,1}+O^{pp}_{s,2}]$ with $O^{pp}_{s,1}$ and $O^{pp}_{s,2}$ given by Eqs (\ref{Eq:T_screened_1})-(\ref{Eq:T_screened_2}), 
\begin{equation}
 \Sigma(11') \approx v_H + iG(13)\left[v_c(13)\delta(1'3)+ W(1 2) O^{pp}_s(35;1'4) G(42)G(25)\right].
 \label{Eqn:S_GWG}
 \end{equation}
Using $v_c(13)=W(13)-W(12)P(25)v_c(53)$ we can rewrite (\ref {Eqn:S_GWG}) as
  \begin{eqnarray}
  \Sigma(11') &\approx&v_H + iG(13)\big[W(12)\delta(1'3)\delta(23)- W(1 2)P(25)v_c(53)\delta(1'3)\nonumber\\
 & +&W(12) O^{pp}_s(35;1'4) G(42)G(25)\big]\nonumber\\
 &=&v_H+iG(13)W(12)\Gamma(31';2),
\end{eqnarray}
where
\begin{equation}
\Gamma(31^\prime;2)=\delta(31^\prime)\delta(21^\prime)-P(25)v_c(53)\delta (31^\prime)+O^{pp}_{s}(35;1^\prime4)G(42)G(25).
\label{Eqn:vertex_from_T_2}
\end{equation}
This is an approximate vertex function that will generate the same self-energy as the pp screened T matrix. In an analogous way one can obtain an approximate vertex function that will yield the same self-energy as the eh screened T matrix: the equation will be the same as (\ref{Eqn:vertex_from_T_2}) with $O^{eh}_{s}$ replacing $O^{pp}_{s}$ on the right-hand side. It is interesting to compare this vertex function with the exact expression derived by Bruneval \textit{et al.} \cite{Bruneval_PRL05} that reads
\begin{equation}
\Gamma(31^\prime;2)=\delta(31^\prime)\delta(21^\prime)+\frac{\delta\Sigma_{xc}(31')}{\delta\rho(5)}P(52).
\label{Eqn:vertex_bruneval}
\end{equation}
One can approximate the second term on the right-hand side of (\ref{Eqn:vertex_bruneval}) as
\begin{eqnarray}
\frac{\delta\Sigma_{xc}(31')}{\delta\rho(5)}P(52)&=&\frac{\delta\left[\Sigma(31')-v_H(3)\delta(31')\right]}{\delta\rho(5)}P(52)\nonumber\\
&=&\frac{\delta\Sigma(31')}{\delta G(45)}\frac{\delta G(45)}{\delta V_{tot}(2)}-P(25)v_c(53)\delta(31')\nonumber\\
&\approx &O_s(35;1'4)G(42)G(25)-P(25)v_c(53)\delta(31')
\label{Eqn:vertex_bruneval_approx}
\end{eqnarray}
with $V_{tot}=U_{ext}+v_H$ as the total classical potential. The last line is an approximation obtained with $\frac{\delta\Sigma}{\delta G}\approx O_s$, where $O_s$ is either the pp or eh screened T matrix, and $\frac{\delta G}{\delta V_{tot}}\approx GG$. Equation (\ref{Eqn:vertex_bruneval_approx}) with (\ref{Eqn:vertex_bruneval}) yields (\ref{Eqn:vertex_from_T_2}). It then becomes clear that the term $-Pv_c$ in Eq.\ (\ref{Eqn:vertex_from_T_2}), that is created by the induced Hartree potential felt by a classical particle, needs to be subtracted since it is already described by the GW approximation for $\Gamma=1$, whereas  the remaining part  is responsible for the induced exchange-correlation potentials felt by a fermion.  We also note that the term $Pv_c$, which comes out in a natural way in our derivation, gives rise to the correlation part of the GW self-energy, $\Sigma^{GW}_c=iGv_cPW=iG(W-v_c)$. It equals the correction done \textit{a posteriori} in the other existent formulations of the screened T matrix to avoid double counting \cite{Aryasetiawan_Tmatrix,Chulkov_PRB,Chulkov_2}. 
\section{Application to the Hubbard molecule\label{Applications}}
In the following the performance of the T matrix as well as the screened T matrix I, and hence of the approximate vertex function (\ref{Eqn:vertex_from_T_2}), is illustrated using the exactly solvable Hubbard molecule at 1/4 and 1/2 filling \cite{Hubbard_63,Hubbard_64} discussed e.g. in Ref.\ \cite{Pina09}. \PR{The Hubbard model is traditionally used to model strongly correlated systems; these are precisely the systems for which GW shows failures. This, together with an exact analytical solution at hand, represents a powerful tool to test the improvements over GW due to the inclusion of $pp$ and $eh$ correlation for low-density systems.}

The Hamiltonian of the Hubbard molecule reads
\begin{equation}
 H = -t \sum_{\substack{i,j=1,2\\i\neq j}}\sum_{\sigma} c^{\dagger}_{i\sigma} c_{j\sigma} + \frac{U}{2}
 \sum_{i=1,2}\sum_{\sigma\sigma^\prime}c^\dagger_{i\sigma}c^\dagger_{i\sigma^\prime}c_{i\sigma^\prime}c_{i\sigma} + \epsilon_0 \sum_{\sigma,i=1,2} n_{i\sigma} + V_0.
 \end{equation}
Here $n_{i\sigma}=c^{\dagger}_{i\sigma}c_{i\sigma}$, $c^{\dagger}_{i\sigma}$ and  $c_{i\sigma}$ are the creation and annihilation operators for an electron at site $i$ with spin $\sigma$, $U$ is the on-site (spin-independent) interaction, $-t$ is the hopping kinetic energy, and $\epsilon_0$ is the orbital energy. The Hamiltonian further contains a potential $V_0$ that can be chosen to fix the zero energy scale. The physics of the Hubbard model arises from the competition between the hopping term, which prefers to delocalize  electrons, and the on-site interaction, which favours localization. The ratio $U/t$ is a measure for the relative contribution of both terms and is the intrinsic, dimensionless coupling constant of the Hubbard model, which will be used in the following.

Projected onto the (orthonormal) site-basis of the Hubbard model the T matrix (\ref{Eqn:T-matrix_pp_1})-(\ref{Eqn:T-matrix_pp_2}) and (\ref{Eqn:T-matrix_eh_1})-(\ref{Eqn:T-matrix_eh_2})  becomes
\begin{equation}
O^{\sigma_1\sigma_2}_{ilkj}(\omega)=-i\delta_{il}\delta_{jk}\left[\bar{O}^{\sigma_1\sigma_2}_{1,ij}(\omega)-\delta_{\sigma_1\sigma_2}\bar{O}^{\sigma_1\sigma_1}_{1,ij}(\omega)\right],
\label{Eqn:uncreened_T_HM}
\end{equation}
with $\bar{O}_1^{\sigma_1\sigma_2}(\omega)=[1+UL^{\sigma_1\sigma_2}_0(\omega)]^{-1}U$, from which
\begin{eqnarray}
\Sigma^{\sigma_1}_{ij}(\omega)&=&-i\int \frac{d\nu}{2\pi}G_{ji}^{\bar{\sigma}_1}(\nu)\bar{O}^{\sigma_1\bar{\sigma}_1}_{ij}(\omega\pm\nu).
\label{Hubbard_SE}
\end{eqnarray}
Here $\bar{\sigma}$ indicates a spin opposite to $\sigma$, the sign $'+'$ refers to the particle-particle contribution for which $L_{0,ij}^{\sigma_1\sigma_2,pp}(\omega)=-i\int \frac{d \omega^\prime}{2\pi} G^{\sigma_1}_{ij}(\omega^\prime)G^{\sigma_2}_{ij}(\omega-\omega^\prime)$, and the sign $'-'$ refers to the electron-hole contribution for which $L_{0,ij}^{\sigma_1\sigma_2,eh}(\omega)=-i\int \frac{d \omega^\prime}{2\pi} G^{\sigma_1}_{ij}(\omega^\prime)G^{\sigma_2}_{ij}(\omega^\prime-\omega)$. More details on the spin-structure of the T matrix can be found in App. \ref{Appendix2}. 
%

The equations for the screened T matrix are more involved: because the screened Coulomb interaction $W$ is nonlocal in space and frequency-dependent (see Ref.\cite{Pina09} and \ref{Eqn:screening}), where a RPA $W$ is given), one has to solve a 4-point equation similar to the dynamical Bethe-Salpeter equation for electron-hole excitations  \cite{Strinati_PRB,Pina_JCP09,Davide_JCP11}. This is beyond the scope of this paper, and of the majority of the applications. Moreover, we observe that when there is no screening in the system (which is the case in the atomic limit  $t\rightarrow 0$ for the model used here) the screened T matrix reduces to the T matrix, and one hence retrieves the same results  as those obtained with the latter. For a finite $t$ we assume that for the model used here the major contribution to the T matrix arises from the on-site screened interaction. This is dominated by the bare interaction $U$, which justifies to take the screened interaction in its static ($\omega=0$) limit. In this case the structure of the screened T matrix is the same as for the T matrix with the onsite screened Coulomb interaction $W=U-\frac{(1+\delta_{N=2})U^2t}{h^2}$, with $h^2=4t^2+2Ut(1+\delta_{N=2})$ and $N$ the total number of electrons in the system (see Ref.\cite{Pina09} and \ref{Eqn:screening})), replacing $U$. We notice that with this approximation, in particular assuming a static $W$, the screened T matrix will not reduce anymore to the T matrix in the atomic limit. This is because in the atomic limit the frequency-dependent part of the screened interaction, i.e. $\frac{(1+\delta_{N=2})U^2t}{(\omega^2-h^2)}$, vanishes if $\omega\neq 0$, but  will reduce to $-\frac{U}{2}$ if $\omega=0$.

\PR{Another simplification that will be adopted in the following is the neglect of self-consistency. In principle the T-matrix approximation requires self-consistency in order to be conserving. The various levels of self-consistency in the T matrix have been addressed by several authors in particular in model systems (see e.g. \cite{Drchal,Verga,Pisarki,Godby_PRB98,Verdozzi_PRL,Verdozzi_PRB}). When the quantity of interest is the spectral function, as in our case,  self-consistency was found to deteriorate spectra \cite{Cini_NC87,C.-O,vonBarth_PRB98,Godby_PRB98,Verdozzi_PRL,Verdozzi_PRB}. In practice, as in the case of the GWA, rather a nonself-consistent "best G" strategy is adopted.} 

  %
 %
\subsection{Hubbard molecule 1/4 filling}
We first consider the Hubbard model with only one electron in the ground state; the ground state is $|\Psi_0\rangle=1/\sqrt{2}(|\uparrow\, 0\rangle+|0\,\uparrow\rangle)$, where the electron has spin up (equivalently the spin-down situation could be chosen). 
\subsubsection{T matrix}
Using the noninteracting Green's function for 1/4 filling of Ref.\ \cite{Pina09} [Eqs (16) and (18)] and Eq.\ (\ref{Hubbard_SE}) the particle-particle contribution to the T matrix yields the exact self-energy, which reads
\begin{eqnarray}
\Sigma^{pp,\uparrow}_{ij}&=&= 0\\
\Sigma^{pp,\downarrow}_{ij}&=&\frac U2\delta_{ij}+\frac{U^2}{8}\left[\frac{1}{\omega-\epsilon_0-\frac U2-3t+i\eta}+\frac{(-1)^{(i-j)}}{\omega-\epsilon_0-\frac U2-t+i\eta}\right].
\end{eqnarray}
This self-energy gives rise to the exact one-particle Green's function and spectral function.

\begin{figure}[htbp]
\begin{center}
\caption{Two-site Hubbard model at 1/4 filling: comparison between the exact spin-down renormalized addition energies $\omega^a/t$ (solid lines) as function of $U/t$ (left panel) and Log(U/t) (right panel) and the results obtained from GW (dashes), particle-particle (solid lines, equal to the exact result), electron-hole (crosses), and 1st iteration T matrix (circles). In the atomic limit the spectral function, i.e. the peak positions and weights, is illustrated on the right-hand side, upon multiplying by $t$ and taking the $t\rightarrow 0$ limit.}
\vspace{4em}
\includegraphics[width=\textwidth]{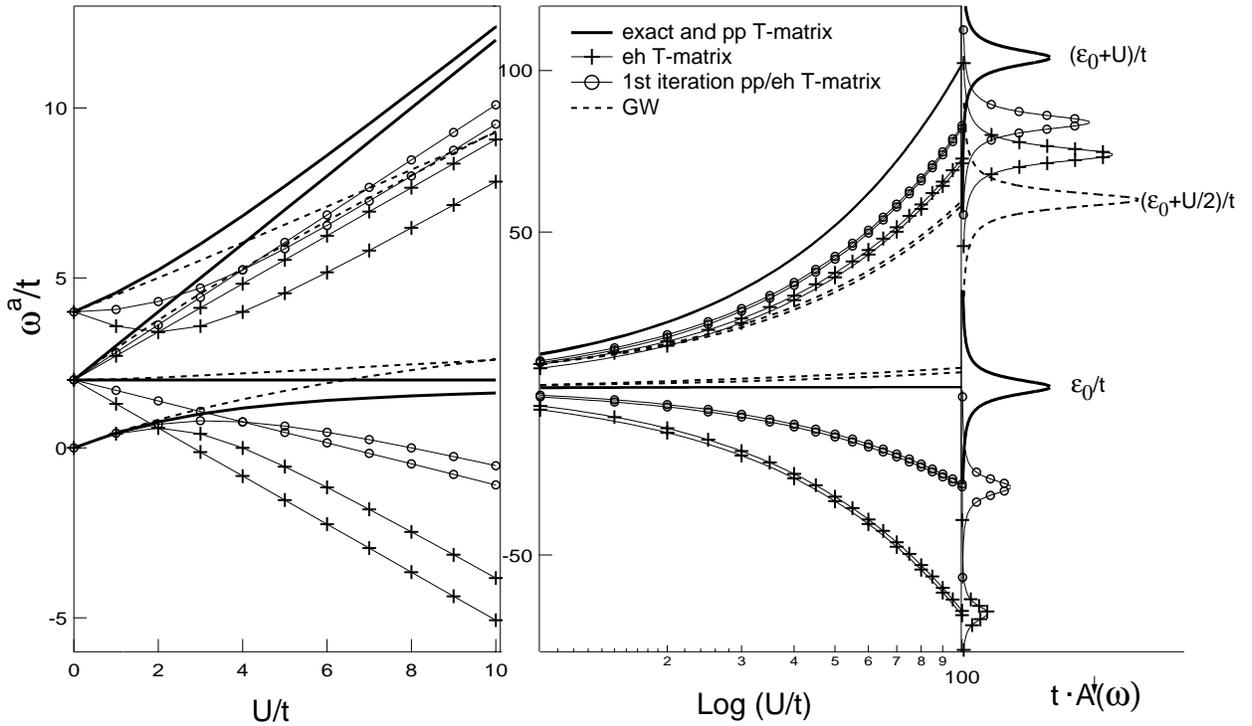}
\label{figure4}
\end{center}
\end{figure}

The eh T matrix also yields the exact result for the spin-up self-energy, but a spin-down self-energy with poles that are shifted by $U$ from the exact ones:
\begin{eqnarray}
\Sigma^{eh,\uparrow}_{ij}&=&0\\
\Sigma^{eh,\downarrow}_{ij}&=&\frac U2\delta_{ij}+\frac{U^2}{8}\left[\frac{1}{\omega-\epsilon_0+\frac U2-3t+i\eta}+\frac{(-1)^{(i-j)}}{\omega-\epsilon_0-t+\frac U2+i\eta}\right].
\end{eqnarray}
The poles of this spin-down self-energy give addition and removal energies in overall bad agreement with the exact result, as shown in Fig.\ \ref{figure4}, where exact, T-matrix, and GW renormalized addition energies $\omega^a/t$ for the spin-down channel are reported $versus$ $U/t$. Particularly interesting is the spectral function at $U/t\rightarrow \infty$: as discussed in Ref.\cite{Pina09}, when $t\rightarrow 0$ (atomic limit) the electron is localized either on one site or the other with the same probability. Therefore two electron peaks, with the same weight $1/2$, appear in the spectral function (see Fig.\ \ref{figure4}), one corresponding to the addition of the second electron on the empty site (peak at $\epsilon_0$) and the other corresponding to the addition on the filled site (peak at $\epsilon_0+U$). The GWA produces only one peak at $\epsilon_0+U/2$ with spectral weight 1. This is due to the interpretation of the charge density as an average charge distribution rather than a probability. The T matrix, instead, "sees" where the electron is, although only the pp T matrix "sees" well. Indeed, the eh T matrix yields two peaks with the correct spectral weight,  but at the wrong position, namely $\epsilon_0+U\sqrt{2}/2$ and $\epsilon_0-U\sqrt{2}/2$.  Therefore, it is clear that in the case of the Hubbard molecule with one electron the particle-particle contribution to the T matrix describes the essential physics. 

Note that the eh T matrix at first iteration performs as the pp T matrix, since $\Sigma^{eh,(1)}=\Sigma^{pp,(1)}$, as we have already shown. In the atomic limit the T matrix at the first iteration shows two peaks in the spin-down spectral function: one located at $\omega=\epsilon_0+U(1-\sqrt{5})/4$ with spectral weight $(1-1/\sqrt{5})/2\approx 0.28$ and the other one located at $\omega=\epsilon_0+U(1+\sqrt{5})/4$ with spectral weight $(1+1/\sqrt{5})/2\approx 0.72$ (see Fig.\ \ref{figure4}). It hence contains the right physics, although the results are still poor. 
\subsubsection{Screened T matrix}
\begin{figure}[htbp]
 \begin{center}
 \caption{Two-site Hubbard model at 1/4 filling: comparison between the exact spin-down renormalized addition energies $\omega^a/t$ (solid lines) as function of $U/t$ (left panel) and Log(U/t) (right panel) and the results obtained from GW (dashes) and particle-particle (solid lines, equal to the exact result) and (approximate) screened T matrix (triangles). In the atomic limit the spectral function, i.e. the peak positions and weights, is illustrated on the right-hand side, upon multiplying by $t$ and taking the $t\rightarrow 0$ limit.}
 \vspace{4em}
 \includegraphics[width=\textwidth]{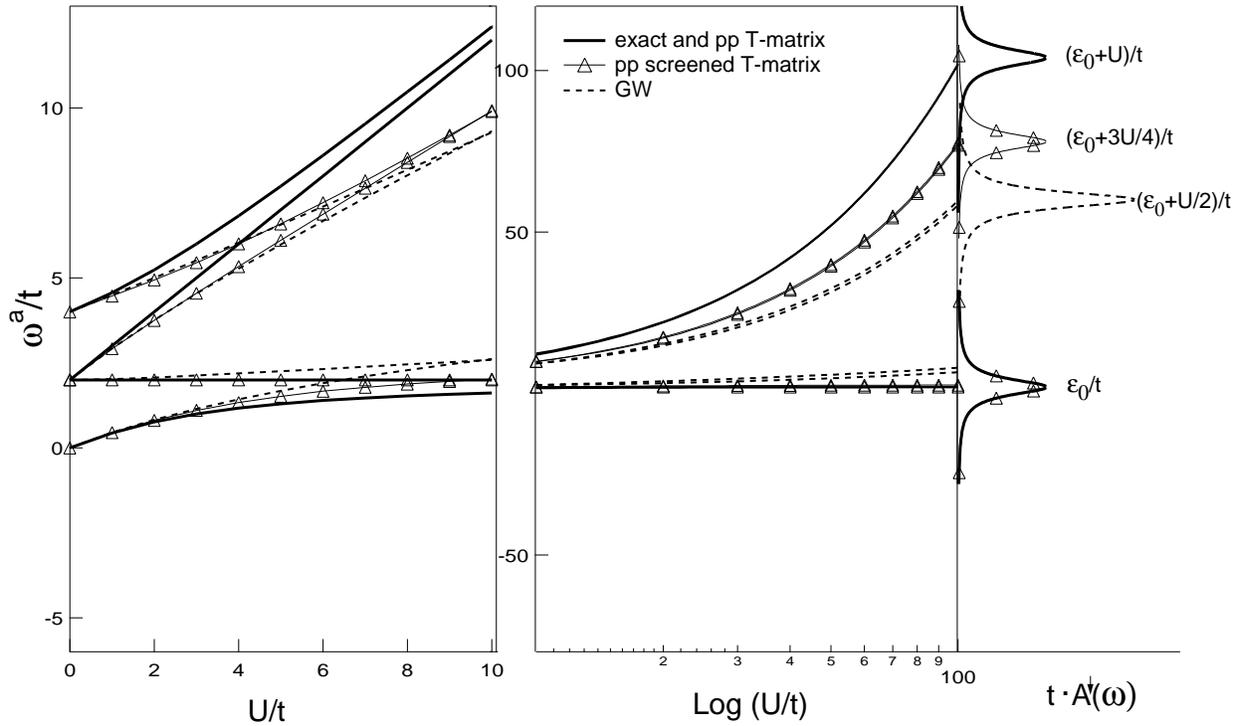}
 \label{figure5}
 \end{center}
 \end{figure}

As illustrated above, the pp T matrix yields the exact result for the Hubbard molecule at 1/4 filling. However, if one is interested in many-electron systems, where screening becomes important, the screened T matrix is more appropriate. It is interesting to check how well the screened T matrix performs in the 1 electron limit. In order to do so we concentrate on the pp screened T matrix only. Using the onsite and instantaneous approximation $W=U-U^2t/h^2$ (with $h^2=4t^2+2Ut$) for the screened Coulomb interaction, the screened pp T matrix yields the self-energy
\begin{eqnarray}
\Sigma^{pp,\uparrow}_{s,ij}&=&= 0\\
\Sigma^{pp,\downarrow}_{s,ij}&=&\frac U2\delta_{ij}+\frac{UW}{8}\left[\frac{1}{\omega-\epsilon_0-\frac W2-3t+i\eta}+\frac{(-1)^{(i-j)}}{\omega-\epsilon_0-\frac W2-t+i\eta}\right],
\end{eqnarray}
which gives rise to the spin-down addition energies reported in Fig.\ \ref{figure5}. The approximate screened T matrix performs in general much better than GW; in particular, in the atomic limit, although it does not reproduce the exact result, it produces the correct number of peaks in the spectral function, unlike the $GW$ approximations which yields only one peak. Therefore,  already in this approximate version the screened T matrix I contains the essential interaction processes. Note that the dynamically screened T matrix I reproduces exactly the atomic limit for the present problem.

The screened T matrix $II$, instead, does not reproduce correctly the atomic limit. Indeed, with $W=v_c$ the screened T matrix II reduces to the $O_2$ component only of the T matrix and this is not sufficient to capture the interactions in the atomic limit. One finds the self-energy
\begin{eqnarray}
\Sigma^{pp,\uparrow}_{ij}(\omega)&=&\Sigma^{eh,\uparrow}_{ij}(\omega)=0\\
\Sigma^{pp,\downarrow}_{ij}(\omega)&=&\Sigma^{eh,\downarrow}_{ij}(\omega)=\frac{U}{2}\delta_{ij},
\end{eqnarray}
which is the same as the one obtained within $GW$.
\subsection{Hubbard molecule 1/2 filling}
We now consider the case with two electrons in the ground state. Technical details are given in App.\ \ref{Hubbard_2e} (see also Ref.\ \cite{Tomczak}). 
\subsubsection{T matrix}
\begin{figure}[htbp]
\begin{center}
\caption{ Two-site Hubbard model at 1/2 filling: comparison between the exact renormalized addition/removal energies $\omega/t$ (solid lines) as function of $U/t$ (left panel) and Log(U/t) (right panel) and the results obtained from GW (dashes), particle-particle (dots), electron-hole (crosses), and 1st iteration  T matrix (solid lines, equal to the exact result). In the atomic limit the spectral function, i.e. the peak positions and weights, is illustrated on the right-hand side, upon multiplying by $t$ and taking the $t\rightarrow 0$ limit.}
\vspace{4em}
\includegraphics[width=\textwidth]{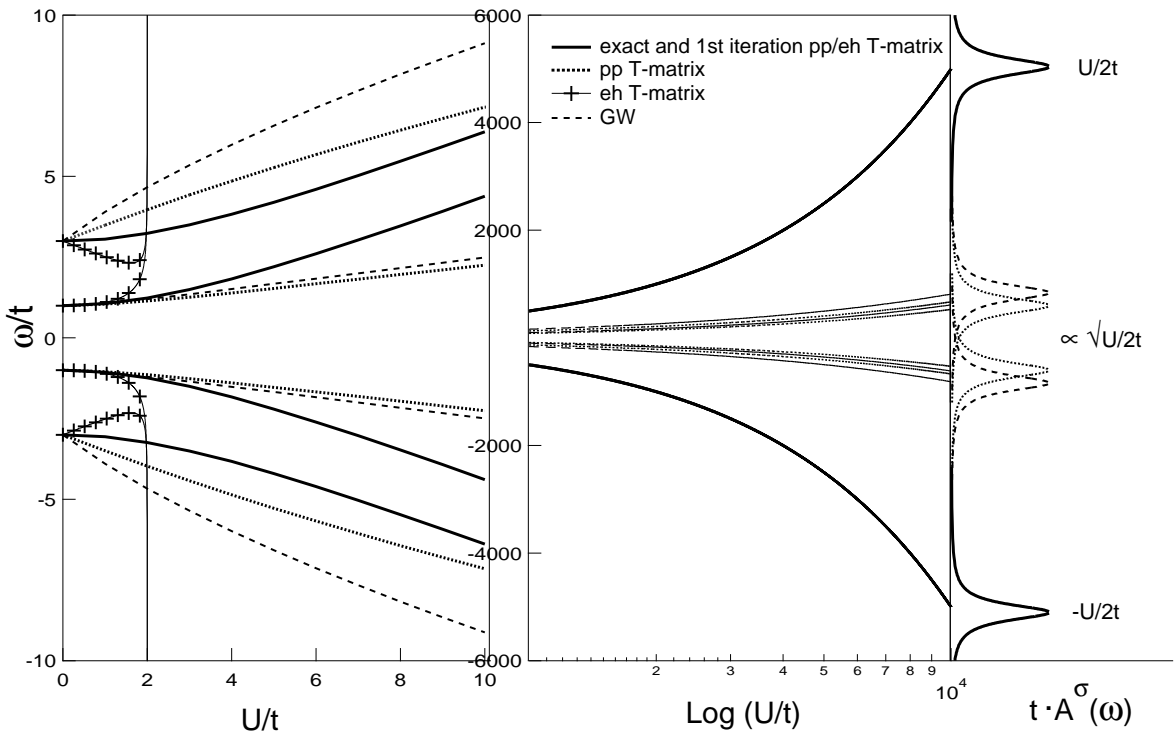}
\label{figure6}
\end{center}
\end{figure}

In this case neither the pp nor the eh T matrix reproduce the exact result, with the self-energy reading as
\begin{eqnarray}
 \Sigma^{pp,\sigma}_{ij}(\omega)&=&\frac{U}{2}\delta_{ij}+\frac{U^2t}{4\bar{h}}\left[\frac{1}{\omega-t-\bar{h}+i\eta}+\frac{(-1)^{(i-j)}}{\omega+t+\bar{h}-i\eta}\right]
 \label{Eqn:sum-diff_selfenergy_pp}
\end{eqnarray}
with $\bar{h}^2=4t^2+2tU$, in the particle-particle T-matrix approximation, and
\begin{eqnarray}
 \Sigma^{eh,\sigma}_{ij}(\omega)&=&\frac{U}{2}\delta_{ij}+\frac{U^2t}{4\bar{h}^\prime}\left[\frac{1}{\omega-t-\bar{h}^\prime+i\eta}+\frac{(-1)^{(i-j)}}{\omega+t+\bar{h}^\prime-i\eta}\right],
 \label{Eqn:sum-diff_selfenergy_eh}
\end{eqnarray}
with $\bar{h}^{\prime 2}=4t^2-2tU$, in the electron-hole T-matrix approximation. The pp T matrix performs rather well over a wide $U/t$ range as one can see in Fig.\ \ref{figure6}, where the renormalized addition/removal energies $\omega/t$ are plotted \textit{versus} $U/t$.  In particular the satellite energies (outer energies) are better described than in GW, in line with previous findings \cite{Aryasetiawan_Tmatrix,Verdozzi_PRL,Verdozzi_PRB}. The energies calculated using the electron-hole T matrix, instead, show divergencies. In the $U/t\rightarrow \infty$ limit, all approximations studied are rather poor.

In the atomic limit there are no double occupancies, therefore the two electrons, one with spin up and the other with spin down, are localized one on one site and the other on the other site with equal probability, i.e. the ground state is the singlet $|\Psi_0\rangle=\frac{1}{\sqrt{2}}(|\uparrow\,\,\downarrow\rangle-|\downarrow\,\,\uparrow\rangle)$. The spectral function thus shows, for each spin, two peaks with the same spectral weight 1/2, one for the removal of an electron (peak at $\epsilon_0$), and one for the addition of a second electron (peak at $ \epsilon_0+U$), as shown in Fig. \ref{figure6}. In App.\ \ref{Hubbard_2e} we show that, using the noninteracting Green's function, GW fails also in the case of 1/2 filling, producing for each spin only one kind of peak at $\epsilon_0 +U/2$ with spectral weight 1/2, both for electron removal and addition. This can again be understood considering that GW treats the charge/spin density as a classical distribution, namely half electron with half spin up and half electron with half spin down on each atom that respond to the additional electron or hole in the atomic limit. We find that the particle-particle T matrix yields the same result as GW in the atomic limit, whereas the electron-hole T matrix shows divergencies. 

{\textit{Why is the pp T matrix exact for one electron in the atomic limit, and not for two electrons?}} To derive the T matrix we used the approximation $\frac{\delta G}{\delta U_{ext}}\approx GG$. In the case of one electron this is not an approximation, but it is the exact time-ordered response, and therefore the (pp) T matrix yields the exact result for one electron. This is not the case for two electrons for which $\frac{\delta G}{\delta U_{ext}}\approx GG$ is a rough approximation, and one needs to include some screening.  The screened T matrix I indeed improves over the T matrix for two electrons even with an approximate RPA screening, as it is shown in the next section; such approximate screening is instead dramatic for one electron and a more accurate screening is needed (as pointed out above the exact screening would yield the T matrix and hence an exact result for one electron). \textit{One should hence use a screened T matrix  with a screened interaction adapted to the system}.

Interestingly the first iteration for both the particle-particle and electron-hole contributions to the T matrix gives the exact results for all $t$. Indeed after the first iteration the
 T matrix reads
\begin{equation}
\bar{O}_{ij}^{\sigma_1\bar{\sigma}_1,(1)}(\omega)=\left[U\delta_{ij}-U^2L_{0,ij}^{\sigma_1\bar{\sigma}_1}(\omega)\right]
\end{equation}
with $L_0$ being
\begin{eqnarray}
L_{0,ij}^{\sigma_1\bar{\sigma}_1,pp}(\omega)&=&\frac{-1}{4}\left[\frac{1}{\omega-2t+i\eta}-\frac{1}{\omega+2t-i\eta} \right],\\
L_{0,ij}^{\sigma_1\bar{\sigma}_1,eh}(\omega)&=&\frac{(-1)^{(i-j)}}{4}\left[\frac{1}{\omega-2t+i\eta}-\frac{1}{\omega+2t-i\eta} \right]
\end{eqnarray}
for the particle-particle and electron-hole contribution, respectively. The self-energy hence becomes
\begin{equation}
\Sigma_{ij}^{pp, \sigma_1, (1)}(\omega)=\Sigma_{ij}^{eh, \sigma_1, (1)}(\omega)=\Sigma_{ij}^{\sigma_1, (1)}(\omega)=\delta_{ij}\frac{U}{2}+\frac{U^2}{8}\left[\frac{1}{\omega-3t+i\eta}+\frac{(-1)^{(i-j)}}{\omega+3t-i\eta}
\right],
\end{equation}
which is the exact one. \PR{This result, however, is peculiar for the Hubbard molecule. Indeed, the first iteration of the T matrix corresponds to the second Born approximation, which has already been explored on bigger Hubbard clusters for different fillings and interactions \cite{Verdozzi_PRL,Verdozzi_PRB}: indeed it does not generate the exact result and in general the T matrix is the most accurate at low densities; only at half-filling the T matrix is not superior to second-Born and this is in line with our findings}.  Within the $GW$ approximation, if one considers $W$ at first iteration ($W^{(1)}_{ij}(\omega)=U\delta_{ij}+U^2\sum_{\sigma}L^{\sigma\sigma,eh}_{0,ij}(\omega)$), the resulting self-energy is not exact, but very close to the exact one, differing only in  the prefactor of the frequency-dependent part that is $U^2/4$ instead of $U^2/8$. The addition and removal energies are thus improved with respect to GW, although the agreement with the exact result worsens with increasing $U$. \PR{It is worth noticing that if one considers also the exchange counterpart in GW, i.e. if one includes not only the Hartree potential but also the exchange self-energy in the self-energy variation $\delta \Sigma/\delta G$  in Eq.\ (\ref{Eqn:Self-energy}), then one obtains a GW-like self-energy with a modified spin-dependent screened interaction $\tilde{W}^{\sigma_1}_{ij}(\omega)=U\delta_{ij}+U\sum_{r,\sigma_r}L^{\sigma_r}_{0,ir}(\omega)\left(1-\delta_{\sigma_r\sigma_i}\right)W_{rj}(\omega)$ that at second order in $U$ produces the exact self-energy. Indeed this is the second Born approximation.}

The exact result obtained with the T matrix at first iteration, however, deteriorates with the second iteration; in this case the T matrix reads
\begin{equation}
\bar{O}_{ij}^{\sigma_1\bar{\sigma}_1,(2)}(\omega)=\bar{O}_{ij}^{\sigma_1\bar{\sigma}_1,(1)}(\omega)+U^3\sum_nL_{0,in}^{\sigma_1\bar{\sigma}_1}(\omega)L_{0,nj}^{\sigma_1\bar{\sigma}_1}(\omega)
\end{equation}
and the two self-energies become
\begin{eqnarray}
\Sigma_{ij}^{pp, \sigma_1, (2)}(\omega)&=&\Sigma_{ij}^{\sigma_1, (1)}(\omega)+\frac{U^3}{16}\left\{ \frac{1}{(\omega-3t+i\eta)^2}-\frac{(-1)^{(i-j)}}{(\omega+3t-i\eta)^2}-\frac{1}{2t}
 \left[\frac{1}{\omega-3t+i\eta}+\frac{(-1)^{(i-j)}}{\omega+3t-i\eta}
\right]\right\},
\label{Eqn:S_O_2_pp}\\
\Sigma_{ij}^{eh,\sigma_1\,(2)}(\omega)&=&\Sigma_{ij}^{\sigma_1, (1)}(\omega)-\frac{U^3}{16}\left\{ \frac{1}{(\omega-3t+i\eta)^2}-\frac{(-1)^{(i-j)}}{(\omega+3t-i\eta)^2}-\frac{1}{2t}
 \left[\frac{1}{\omega-3t+i\eta}+\frac{(-1)^{(i-j)}}{\omega+3t-i\eta}
\right]\right\}.
\label{Eqn:S_O_2_eh}
\end{eqnarray}
Combining the two interaction channels by adding Eqs (\ref{Eqn:S_O_2_pp}) and (\ref{Eqn:S_O_2_eh}), the second terms on the right-hand side of the two equations cancel each other, thus restoring the exact result if $\frac{1}{2}\left(\Sigma^{pp,(2)}+\Sigma^{eh,(2)}\right)$ is taken. We have already shown in Eq.\ (\ref{Eqn:Functional_derivative_2}) that the sum $\frac{1}{2}\left(\Sigma^{pp,(2)}+\Sigma^{eh,(2)}\right)$ takes into account some of the terms which would appear in the self-energy if also the functional derivative $\delta O/\delta G$ were considered. This might justify the exact result that is obtained by taking $\frac{1}{2}\left(\Sigma^{pp,(2)}+\Sigma^{eh,(2)}\right)$. Also the sum $\left(\Sigma^{pp,(2)}+\Sigma^{eh,(2)}\right)$ takes into account some of the terms arising from the functional derivative (see Eq.\ (\ref{Eqn:Functional_derivative_2})); however it does not give the exact result. In other words it seems more important to take into account the term $-T_0$ than $(T_1+T_2)/2$ (see Eqs (\ref{Eqn:Functional_derivative_1}-\ref{Eqn:Functional_derivative_2})) at least in the present problem. However with the third iteration the fourth order terms in the pp and eh T matrix self-energies are the same and they would not cancel each other if the sum $\frac{1}{2}\left(\Sigma^{pp,(2)}+\Sigma^{eh,(2)}\right)$ is taken. Instead with the forth iteration the fifth order terms in the pp and eh T matrix self-energies would cancel each other. \PR{In general for the present problem the pp and eh T matrix self-energies show, starting from second order, the same even-order terms and opposite odd-order terms, as one can verify Taylor expanding the frequency-dependent part of the pp and eh self-energies (Eqs (\ref{Eqn:sum-diff_selfenergy_pp}) and (\ref{Eqn:sum-diff_selfenergy_eh})) for small $U$. The same holds for the Hubbard model at 1/4 filling.} Therefore summing the two contributions will not give the exact result. \PR{Even adding the GW self-energy terms and its exchange counterparts, in the spirit of the FLEX approximation, will not produce the exact result. 

These findings show that there is no an ultimate way to combine diagrams, and this is of clear relevance for realistic studies where several attempts to combine pp and eh channels have been done (see e.g. \cite{Liebsch,Katsnelson_JPCM,Katsnelson_EPJB,Chulkov_PRB}).  A possibility is to use the screened T matrix we introduced in Sec.\  \ref{Screened_T_ok}, which produces, at least for the studied problem, results overall better than GW and T matrix, as shown in the following.}

\subsubsection{Screened T matrix}
\begin{figure}[htbp]
 \begin{center}
 \caption{Two-site Hubbard model at 1/2 filling: comparison between the exact renormalized addition/removal energies $\omega/t$ (solid lines) as function of $U/t$ (left panel) and Log(U/t) (right panel) and the results obtained from GW (dashes) and particle-particle (dots) and screened T matrix (triangles). In the atomic limit the spectral function, i.e. the peak positions and weights, is illustrated on the right-hand side, upon multiplying by $t$ and taking the $t\rightarrow 0$ limit.}
 \vspace{4em}
 \includegraphics[width=\textwidth]{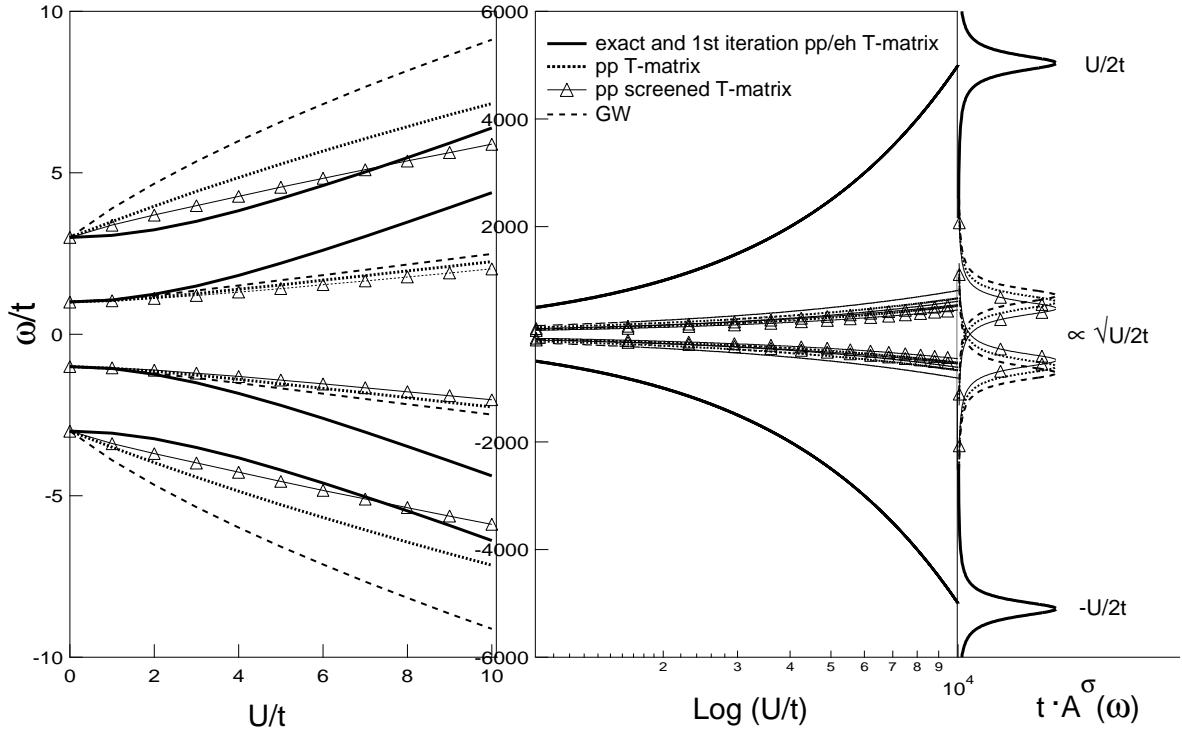}
 \label{figure7}
 \end{center}
 \end{figure}

The pp screened T matrix with the approximate onsite static Coulomb interaction $W=U-2U^2t/h^2$ (with $h^2=4t^2+4Ut$) leads to the self-energy
\begin{eqnarray}
 \Sigma^{pp,\sigma}_{s,ij}(\omega)&=&\frac{U}{2}\delta_{ij}+\frac{UWt}{4\tilde{h}}\left[\frac{1}{\omega-t-\tilde{h}+i\eta}+\frac{(-1)^{(i-j)}}{\omega+t+\tilde{h}-i\eta}\right]\label{Eqn:screened_selfenergy_pp}
\end{eqnarray}
with $\tilde{h}^2=4t^2+2tW$. The resulting renormalized addition and removal energies $\omega/t$ are plotted in Fig.\ \ref{figure7} $versus$ $U/t$, and compared with the exact, pp T matrix, and GW results. This approximate pp screened T matrix is overall superior to the GW and the T matrix in the selected $U/t$ range in the left panel of Fig.\ \ref{figure7}. In the limit $U/t\rightarrow\infty$ the results get corrupted. As a consequence  the screened T matrix performs  as poor as GW and the T matrix. 
\\

\PR{\
Our findings suggest that the screened T-matrix approximation (or equivalently $GW\Gamma$, with $\Gamma$ obtained from the screened T matrix ) is expected to describe properly also larger, more dense systems. This is in agreement with the idea behind the screened T matrix to combine T matrix and GW and to take advantage of the strength of both approaches.
For short-range interactions, where screening is not important, the screened T matrix reduces to the T matrix, which is suitable for treating short-range correlation. For long-range interaction, where, instead, screening is important, we find that the screened T matrix behaves more like $GW$ (in its first iteration, indeed, it gives GW and SOSEX, which is actually already used to improve $GW$), which is capable of taking into account long-range correlation. Therefore the screened T matrix is able to capture the physics of systems with effective short-range interactions as well as of systems with effective long-range interactions. 

For practical calculations the equations become very involved. However, the Dyson equation for the T matrix is similar to the Bethe-Salpeter equation for the eh excitations, therefore one might use similar strategies to solve it. For example one may consider the static approximation to the screened Coulomb interaction, as usually used for the BSE and as adopted for the screened T matrix of Ref.\  \cite{Aryasetiawan_Tmatrix}. For short-range interactions one can consider, together with the static approximation for $W$, also a local approximation for the four-point kernel $GG$ which was proposed in Ref.\ \cite{Karlsson_PRB00} for a similar screened T matrix and validated by several applications \cite{Karlsson_PRB00,Chulkov_PRB,Chulkov_PRL}; this simplifies a lot the calculation of the screened T matrix, which becomes a two-point quantity similar to the screened Coulomb potential $W$, and opens the way for wide applications.}%

\section{Conclusions\label{Conclusions}}
In this paper we have given an alternative derivation of the T-matrix approximation to the self-energy starting from exact equations. This allowed us to: i) link the T matrix to Hedin's equations; ii) understand the origin of the electron-hole (eh) and the particle-particle (pp) T matrix; iii) derive a screened T matrix; iv) translate the physical content of the T matrix into a vertex correction; v) put the second-order screened exchange (SOSEX) on the same level as the T matrix.

We applied the T matrix to the exactly solvable Hubbard molecule at 1/4 and 1/2 filling and we studied its performance with increasing ratio $U/t$. We found that the particle-particle T matrix gives the exact removal and addition energies for 1/4 filling. The electron-hole T matrix, instead, performs badly. 

In the case of 1/2 filling the pp  performs in general better than $GW$, in particular in describing the satellite position, whereas the eh T matrix show divergencies for $U/t>2$. In the atomic limit $t\rightarrow 0$ both GW and T matrix are very poor. In their first iteration both particle-particle and electron-hole T matrix produce the exact result. This result gets corrupted with further iterations; however pp and eh self-energies have the same even-order terms and opposite odd-order terms (except the first order term which is the same) at least in the model analyzed here. This means that one gets the exact self-energy if  the sum $\frac{1}{2}\left(\Sigma^{pp,(2)}+\Sigma^{eh,(2)}\right)$ at the second iteration is taken, and one retains only the even-order terms if the eh an pp self-energies are taken at infinite order. This can be of relevance in realistic calculations where eh and pp channels are combined together for improving the results. 

We have also studied the performance of the pp screened T matrix. The screened T matrix I (see text), in which Hartree and exchange terms are treated on an equal footing, reduces to the T matrix in the atomic limit when a dynamically screened interaction is used; even with an approximate $W$ it is better than GW at 1/4 filling, whereas at 1/2 filling it is overall superior to both the pp T matrix and GW over a wide $U/t$ range. This means that the vertex corrections derived from this version of the  pp screened T matrix can visibly improve over GW. The screened T matrix II (see text), in which only exchange-like terms are included, reduces to only one part of the T matrix in the atomic limit and this is not sufficient to describe exactly this limit at 1/4 filling. We show that this version of the screened T matrix corresponds at first iteration to the $GW\Gamma^{(1)}$ approximation, which hence is also not appropriate to treat the atomic limit.  \PR{Our illustration of the different T-matrix approximations on the Hubbard molecule suggests that the screened T matrix I is a promising approximation also for realistic systems, since it combines on an equal footing on one hand the physics of the T-matrix approximation, which properly describes short-range interaction, and on the other hand the physics of $GW$, which is more appropriate for long-range interaction. }
\begin{acknowledgments}
We are grateful for the support by ANR (Project No. NT09-610745 ). Financial support due to the French-German Gay-Lussac Humboldt Award and the kind hospitality of the \'Ecole Polytechnique is acknowledged by F.B..
\end{acknowledgments}
\appendix
\section{Time structure of the T matrix\label{Appendix1}}
First we examine the time structure of the T matrix (\ref{Eqn:T-matrix_pp_1})-(\ref{Eqn:T-matrix_pp_2}) and (\ref{Eqn:T-matrix_eh_1})-(\ref{Eqn:T-matrix_eh_2}).
\subsection{particle-particle T matrix}
We start from Eqs (\ref{Eqn:T-matrix_pp_1})-(\ref{Eqn:T-matrix_pp_2}). In the following the indices will refer to the time only. We can define $O^{pp}_1(12; 1'4) := -i\delta(12)\bar{O}^{pp}_1(11; 1'4)$ which leads to
\begin{equation}
\label{eq:o1-time-2}
 \bar O^{pp}_1(1-1';1-4) = v_c\delta(11')\delta(41^+)+iv_c G(1-3)\bar O^{pp}_1(3-1';3-4) G(1^+-3).
\end{equation}
Here the correct order of the field operators is explicitly ensured by the infinitesimal larger $1^+=1+\eta>1$. The two Green's functions in the product have the same time order, contrary to the usual electron-hole response function. This leads to a particle-particle function $L^{pp}_0(1-3):=-iG(1-3)G(1^+-3)$. We then get in frequency space
\begin{equation}
\label{eq:o1-time-omega}
 \bar O^{pp}_1(\omega;\omega') = v_c-v_cL^{pp}_0(\omega+\omega')\bar O^{pp}_1(\omega;\omega'),
\end{equation}
with $L^{pp}_0(\omega)=-i\int\frac{d\omega'}{2\pi}G(\omega')G(\omega-\omega')e^{i\omega'\eta}$.
This finally implies that $\bar O^{pp}_1$ depends only on the sum of frequencies:
\begin{equation}
\label{eq:o1-time-omega-diff}
 \bar O^{pp}_1(\omega+\omega') = v_c-v_cL^{pp}_0(\omega+\omega')\bar O^{pp}_1(\omega+\omega').
\end{equation}
This is a Dyson-like equation similar to the screening equation in the GW approximation. The time structure of the self-energy becomes
\begin{equation}
 \Sigma_1(1-1')=G(4-2^+)O^{pp}_1(12;1'4)=-iG(4-1^+)\bar O^{pp}_1(1-1';1-4)
\end{equation}
and hence
\begin{equation}
 \Sigma^{pp}_1(\omega)=-i\int \frac{d\omega'}{2\pi}G(\omega')\bar O^{pp}_1(\omega;\omega')e^{i\omega'\eta}=-i\int \frac{d\omega'}{2\pi}G(\omega')\bar O^{pp}_1(\omega+\omega')e^{i\omega'\eta}.
\end{equation}
Again, this is very close to the structure of GW; simply, one has a ``particle-particle-screened'' interaction $-\bar O^{pp}_1$. One can verify that the time structure of $O^{pp}_2$ is the same as for $O^{pp}_1$.
\subsection{electron-hole T matrix}
In case of the electron-hole T matrix (\ref{Eqn:T-matrix_eh_1})-(\ref{Eqn:T-matrix_eh_2}) one can do similar steps as above and arrives at
\begin{equation}
 \bar O^{eh}_1(1-5;1-1') = v_c\delta(11')\delta(1^+5)-v_c L^{eh}_0(1-3)\bar O^{eh}_1(3-5;3-1'),
\end{equation}
where we defined the electron-hole function$L^{eh}_0(1-3):=-iG(1-3)G(3-1^+)$. In Fourier space one gets
\begin{equation}
\bar{O}^{eh}_1(\omega+\omega')=v_c-v_c L^{eh}_0(\omega+\omega')\bar{O}^{eh}_1(\omega+\omega'),
\end{equation}
with $L^{eh}_0(\omega)=-i\int\frac{d\omega'}{2\pi}G(\omega')G(\omega'-\omega)e^{i\omega'\eta}$.

The time structure of the self-energy becomes
\begin{equation}
 \Sigma_1(1-1')=G(2-5^+)O^{eh}_1(15;1'2)=-iG(1-5^+)\bar O^{eh}_1(1-5;1-1')
\end{equation}
hence
\begin{equation}
\Sigma^{eh}_1(\omega)=-i\int \frac{d\omega'}{2\pi} G(\omega')\bar{O}^{eh}_1(\omega';\omega)e^{i\omega'\eta}=-i\int \frac{d\omega'}{2\pi} G(\omega')\bar{O}^{eh}_1(\omega-\omega')e^{i\omega'\eta}.
\end{equation}

For $O^{eh}_2$ one can proceed in a similar way as above.

\section{Spin structure of the T matrix\label{Appendix2}}
We now schematize the spin structure of the T matrix. Both for pp and eh T matrix one has, in the collinear limit, $\Sigma(\sigma)=G(\sigma_2)O(\sigma\sigma_2;\sigma\sigma_2)$, with
\begin{equation}
\label{eq:o1-spin}
 O_1(\sigma\sigma_2;\sigma\sigma_2) = -iv_c+ iv_cG(\sigma) O_1(\sigma\sigma_2;\sigma\sigma_2) G(\sigma_2),
\end{equation}
and 
\begin{equation}
\label{eq:o2-spin}
O_2(\sigma\sigma_2;\sigma\sigma_2) = iv_c\delta_{\sigma\sigma_2}+iv_c G(\sigma) O_2(\sigma\sigma_2;\sigma\sigma_2) G(\sigma_2).
\end{equation}
Note that, unlike the screened interaction $W$ used in the $GW$ approximation, the T matrix is spin-dependent.  
\section{Hubbard molecule at half filling\label{Hubbard_2e}}
The starting point is the following Hubbard Hamiltonian
\begin{equation}
 H = -t \sum_{\substack{i,j=1,2\\i\neq j}}\sum_{\sigma} c^{\dagger}_{i\sigma} c_{j\sigma} + \frac{U}{2}
 \sum_{i=1,2}\sum_{\sigma\sigma^\prime}c^\dagger_{i\sigma}c^\dagger_{i\sigma^\prime}c_{i\sigma^\prime}c_{i\sigma} + \epsilon_0 \sum_{\sigma,i=1,2} n_{i\sigma} + V_0.
\end{equation}
Here $n_{i\sigma}=c^{\dagger}_{i\sigma}c_{i\sigma}$, $c^{\dagger}_{i\sigma}$ and  $c_{i\sigma}$ are the creation and annihilation operators for an electron at site $i$ with spin $\sigma$, $U$ is the on-site (spin-independent) interaction, $-t$ is the hopping kinetic energy, and $\epsilon_0$ is the orbital energy. The Hamiltonian further contains a potential $V_0$ that can be chosen to fix the zero energy scale. The eigenstates of the system will be linear combinations of Slater determinants, which are denoted by the kets $|1\,\,\, 2\rangle$, with occupations of the sites 1, 2 given by 0, $\uparrow$, $\downarrow$, $\uparrow\downarrow$. We choose $\epsilon_0=-\frac{U}{2}$ and $V_0=\frac{U}{2}N$, where $N$ is the total number of electrons in the system, i.e. 2 in our case.  This choice is particular convenient since the obtained Hamiltonian,
\begin{equation}
 H = -t \sum_{\substack{i,j=1,2\\i\neq j}}\sum_{\sigma} c^{\dagger}_{i\sigma} c_{j\sigma} + \frac{U}{2}
 \sum_{i=1,2}\sum_{\sigma\sigma^\prime}c^\dagger_{i\sigma}c^\dagger_{i\sigma^\prime}c_{i\sigma^\prime}c_{i\sigma} -\frac{U}{2} \sum_{\sigma,i=1,2} n_{i\sigma} + U,
\end{equation}
has high symmetry (particle-hole symmetry), as we shall see below (see also \cite{Tomczak}). Using this Hamiltonian we can calculate the exact one particle Green's function.
%
%
\subsection{Exact solution}
The main exact quantities we are interested in are
\begin{eqnarray}
G^{\sigma}_{ij}(\omega)&=&\frac{(-1)^{(i-j)}}{2a^2}\left[ \frac{(1+\frac{4t}{(c-U)})^2}{\omega-(c/2-t)+i\eta}+ \frac{(-1)^{(i-j)}(1-\frac{4t}{(c-U)})^2}{\omega-(c/2+t)+i\eta}\right]\nonumber\\
&+&\frac{1}{2a^2}\left[ \frac{(1+\frac{4t}{(c-U)})^2}{\omega+(c/2-t)-i\eta}+ \frac{(-1)^{(i-j)}(1-\frac{4t}{(c-U)})^2}{\omega+(c/2+t)-i\eta}\right],
\label{Eqn:G_11}
\end{eqnarray}
\begin{eqnarray}
G^{0,\sigma}_{ij}(\omega)&=&\frac{(-1)^{(i-j)}}{2}\left[ \frac{1}{\omega-t+i\eta}+ \frac{(-1)^{(i-j)}}{\omega+t-i\eta}\right],
\label{Eqn:G_0-2e}
\end{eqnarray}
\begin{eqnarray}
\Sigma^{\sigma}_{ij}(\omega)&=&\frac{U}{2}\delta_{ij}+\frac{U^2}{8}\left[ \frac{1}{\omega-3t+i\eta}+ \frac{(-1)^{(i-j)}}{\omega+3t-i\eta}\right],
\label{Eqn:SE_exact_HM}
\end{eqnarray}
with $c^2=16t^2+U^2$ and $a^2=2\left(\frac{16t^2}{(c-U)^2}+1\right)$. Note that the symmetry of the system is such that $G^\uparrow_{11}=G^{\downarrow}_{11}=G^\uparrow_{22}=G^{\downarrow}_{22}$ and $G^\uparrow_{12}=G^{\downarrow}_{12}=G^\uparrow_{21}=G^{\downarrow}_{21}$ and similarly for the self-energy.%
\subsection{GW solution}
Also here we give the main results
\begin{eqnarray}
P^{\sigma\sigma}_{ij}(\omega)&=&\frac{(-1)^{(i-j)}}{4}\left[\frac{1}{\omega-2t+i\eta}-\frac{1}{\omega+2t-i\eta}\right]
\label{Eqn:polarizability}
\end{eqnarray}
\begin{eqnarray}
W_{ij}(\omega)&=&U\delta_{ij}+(-1)^{(i-j)}\frac{U^2t}{h}\left[\frac{1}{\omega-h+i\eta}-\frac{1}{\omega+h-i\eta}\right]
\label{Eqn:screening}
\end{eqnarray}
\begin{eqnarray}
\Sigma^{\sigma}_{ij}(\omega)&=&\frac{U}{2}\delta_{ij}+\frac{U^2t}{2h}\left[\frac{1}{\omega-(t+h)+i\eta}+\frac{(-1)^{(i-j)}}{\omega+(t+h)-i\eta}\right],
\end{eqnarray}
where $h^2=4t^2+4tU$. Note that only the component $P^{\sigma\sigma}$ is given in (\ref{Eqn:polarizability}) (and not the full $P^{\sigma\sigma'}$), since only this is needed to calculate $W$.

The poles of the one-particle Green's function can be calculated using
\begin{eqnarray}
det[G^{-1}]=det[G_0^{-1}-\Sigma]=det \left(\begin{array}{cc} \omega-\Sigma_{11}& t-\Sigma_{12} \\ 
t-\Sigma_{12}& \omega-\Sigma_{11}\\ \end{array}\right)
\label{Eqn:General_poles}
\end{eqnarray}
from where we get
\begin{equation}
\omega= t+(\Sigma_{11}-\Sigma_{12}),\,\,\omega= -t+(\Sigma_{11}+\Sigma_{12})
\end{equation}
which is general, i.e. we can apply for any approximation to the self-energy. In the case of GW, we get the following poles
\begin{eqnarray}
\omega_{1,2}&=&\frac{\frac{U}{2}-h\pm\sqrt{(h+\frac{U}{2}+2t)^2+\frac{4U^2t}{h}}}{2}
\label{Eqn:energies_1-2}\\
\omega_{3,4}&=&\frac{\frac{U}{2}+h\pm\sqrt{(h-\frac{U}{2}+2t)^2+\frac{4U^2t}{h}}}{2}.
\label{Eqn:energies_3-4}
\end{eqnarray}
We note that for $U\neq0$ the particle-hole symmetry is lost due to the lack of self-consistency using the $G_0W_0$ approximation \cite{Pina09}. This symmetry can be enforced by absorbing the static part of the self-energy ($U/2$) into the chemical potential; this ultimately corresponds to dropping the terms U/2 in (\ref{Eqn:energies_1-2})-(\ref{Eqn:energies_3-4}).
%

\bibliographystyle{apsrev4-1}
\newpage
\newpage
\newpage

\newpage
 \newpage

\newpage
 \newpage

\end{document}